\documentstyle[12pt,epsf]{article}

\renewcommand{\thefootnote}{\fnsymbol{footnote}}
\newcommand{\ket}[1]{\vert\,{#1}\rangle}
\newcommand{\VEV}[1]{\left\langle{#1}\right\rangle}
\newcommand{\zetahalf}{{\zeta\over 2}}
\def\ru1{\rule[-0.4truecm]{0mm}{1truecm}}
\begin{document}
\begin{flushright}
SLAC-PUB-8472 \\
September 2000
\end{flushright}
\bigskip\bigskip
{\centerline{\Large
\bf Light-Cone Wavefunction Representation}}
\medskip
{\centerline{\Large
\bf of Deeply Virtual Compton Scattering
\footnote{\baselineskip=13pt
Work partially supported by the Department of Energy, contract
DE--AC03--76SF00515.
}}}
\vspace{22pt}
\centerline{ \bf Stanley J.  Brodsky\,$^a$, Markus
Diehl\,$^a$\footnote{Supported by the Feodor Lynen Program of the
Alexander von Humboldt Foundation.}, and Dae Sung Hwang\,$^b$}
\vspace{8pt}
{\centerline{$^a$ Stanford Linear Accelerator Center,}}
{\centerline{Stanford University, Stanford, California 94309, USA}}
\centerline{e-mail: sjbth@slac.stanford.edu, markus.diehl@desy.de}
\vspace{8pt}
{\centerline{$^{b}$ Department of Physics, Sejong University, Seoul
143--747, Korea}}
\centerline{e-mail: dshwang@kunja.sejong.ac.kr}
%
%
\vspace{10pt}
\begin{center}
{\large \bf Abstract}
\end{center}
We give a complete representation of virtual Compton scattering
$\gamma^* p \to \gamma p$ at large initial photon virtuality $Q^2$ and
small momentum transfer squared $t$ in terms of the light-cone
wavefunctions of the target proton. We verify the identities between
the skewed parton distributions $H(x,\zeta,t)$ and $E(x,\zeta,t)$
which appear in deeply virtual Compton scattering and the
corresponding integrands of the Dirac and Pauli form factors $F_1(t)$
and $F_2(t)$ and the gravitational form factors $A_{q}(t)$ and
$B_{q}(t)$ for each quark and anti-quark constituent.  We illustrate
the general formalism for the case of deeply virtual Compton
scattering on the quantum fluctuations of a fermion in quantum
electrodynamics at one loop.
\vfill
\centerline{
}
\vfill
\centerline{Submitted to Nuclear Physics B.}
\vfill
\newpage
\setcounter{footnote}{0}
\renewcommand{\thefootnote}{\arabic{footnote}}

\section{Introduction}

\begin{figure}[htb]
\begin{center}
\leavevmode
\epsfbox{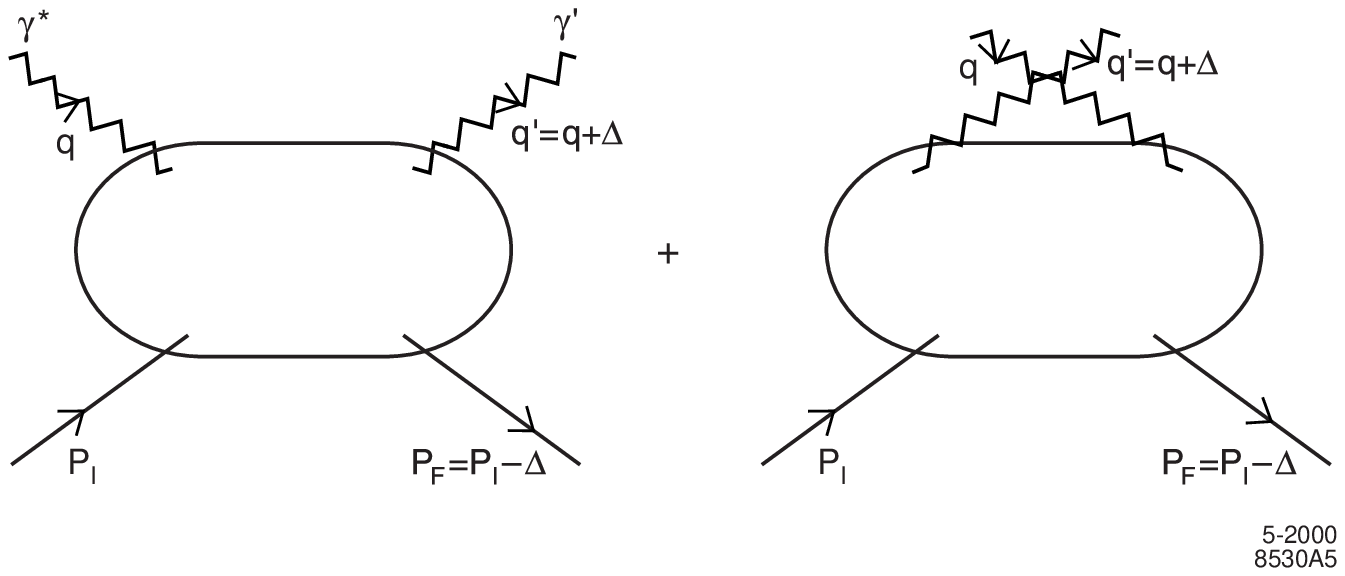}
\end{center}
\caption[*]{The virtual Compton amplitude $\gamma^*(q) + p(P) \to
\gamma(q') + p(P')$.}
\label{fig:1}
\end{figure}

Virtual Compton scattering $\gamma^* p \to \gamma p$ (see
Fig.~\ref{fig:1}) has extraordinary sensitivity to fundamental
features of the proton's structure. Particular interest has been
raised by the description of this process in the limit of large
initial photon virtuality $Q^2 = -q^2$
\cite{Muller:1994fv,Ji:1997ek,Radyushkin:1997ki,Ji:1998pc,Blumlein:1999sc}.
Even though the final state photon is on-shell, one finds that the
deeply virtual process probes the elementary quark structure of the
proton near the light-cone as an effective local current, or in other
words, that QCD factorization
applies~\cite{Radyushkin:1997ki,Ji:1998xh,Collins:1999be}

In contrast to deep inelastic scattering, which measures only the
absorptive part of the forward virtual Compton amplitude,
${\mathrm{Im}}\, {\cal T}_{\gamma^* p \to \gamma^* p}$, deeply virtual
Compton scattering allows the measurement of the detailed momentum and
spin structure of proton matrix elements for general squared momentum
transfer $t=(P-P')^2$.  In addition, the interference of the
amplitudes for virtual Compton scattering and the Bethe-Heitler
process, where the photon is emitted from the lepton line, leads to an
electron-positron asymmetry in the ${e^\pm p \to e^\pm p \gamma}$
cross section which is proportional to the real part of the Compton
amplitude \cite{Brodsky:1972zh}. The imaginary part can be accessed
through various spin asymmetries \cite{Kroll:1996pv}.  The deeply
virtual Compton amplitude $\gamma^* p \to \gamma p$ is related by
crossing to another important process, $\gamma^* \gamma \to$
\textit{hadron pairs} at fixed invariant mass, which can be measured
in electron-photon collisions \cite{Muller:1994fv,Diehl:2000uv}.

\vspace{.5cm}

To leading order in $1/Q$, the deeply virtual Compton scattering
amplitude factorizes as the convolution in $x$ of the amplitude for
hard Compton scattering on a quark line with skewed parton
distributions $H(x,\zeta,t),$ $ E(x,\zeta,t)$, $\widetilde
H(x,\zeta,t),$ and $\widetilde E(x,\zeta,t)$ of the target proton.
Here $x$ is the light-cone momentum fraction of the struck quark, and
$\zeta= Q^2/(2 P\cdot q)$ plays the role of the Bjorken variable known
from deep inelastic scattering. One can also interpret these
distributions in terms of virtual \emph{quark-proton} scattering
amplitudes as defined in the covariant parton model
\cite{Brodsky:1972zh,Landshoff:1971ff,Diehl:1998sm}.

There are remarkable sum rules connecting $H(x,\zeta,t)$ and
$E(x,\zeta,t)$ with the corresponding helicity conserving and helicity
flip electromagnetic form factors $F_1(t)$ and $F_2(t)$ and
gravitational form factors $A_{q}(t)$ and $B_{q}(t)$ for each quark
and anti-quark constituent \cite{Ji:1997ek}.  For example, the
gravitational form factors are given by
\begin{equation}
\int_0^1 {{\rm d}x \over 1 - \zetahalf}\
{x - {\zeta\over 2} \over 1 - \zetahalf}\
[H(x,\zeta,t)  + E(x,\zeta,t)] \ =\ A_{q}(t) + B_{q}(t)
\ .
\label{com3z}\\
\nonumber
\end{equation}
Thus deeply virtual Compton scattering is related to the quark
contribution to the form factors of a proton scattering in a
gravitational field. The total anomalous gravito-magnetic
moment $B(t = 0)$ vanishes identically when summed over all
constituents \cite{Brodsky:2000ii}. In the present work the close
connection between skewed parton distributions and hadronic form
factors will become apparent. To emphasize this relationship, we will
refer to $H, E, \widetilde H$ and $\widetilde E $ as ``generalized
Compton form factors''.

It has long been known that the conventional parton distributions
which describe deep inelastic scattering can be represented in terms
of the squared light-cone Fock-state wavefunction of the proton target
\cite{Lepage:1980fj}. This representation reflects the fact that
parton distributions can be understood as probability densities. In
contrast, virtual Compton scattering always involves non-zero momentum
transfer, and a probabilistic interpretation of skewed parton
distributions is not possible. However, these distributions can still
be constructed from specific overlap integrals of the proton
wavefunctions. They can in fact be regarded as \emph{interference
terms} between wavefunctions for different parton configurations,
containing information on the proton structure that is not accessible
at the level of probability densities. This overlap representation is
the focus of the present work.

There are three distinct integration regions in $x$. In the domain
where $\zeta < x < 1$, the generalized form factors $H$, $E$,
$\widetilde H$ and $\widetilde E$ correspond to the situation where
one removes a quark from the initial proton wavefunction at light-cone
momentum fraction $x=k^+/P^+$ and transverse momentum ${\vec k_\perp}$
and re-inserts it into the final-state wavefunction of the proton with
the same chirality, but with light-cone momentum fraction $x- \zeta$
and transverse momentum ${\vec k_\perp} -{\vec \Delta_\perp}$. The
domain $\zeta-1 < x < 0$ corresponds to removing an antiquark with
momentum fraction $\zeta-x $ and re-inserting it with momentum $-x$,
both momentum fractions being positive as they must. In the remaining
integration domain, $0< x < \zeta$, the photons scatter off of a
virtual quark-antiquark pair in the initial proton wavefunction: the
quark of the pair has light-cone momentum fraction $x$ and transverse
momentum ${\vec k_\perp}$, whereas the anti-quark has light-cone
momentum fraction $\zeta-x$ and transverse momentum ${\vec
  \Delta_\perp} - {\vec k_\perp}$. This domain is unique to skewed
parton distributions and does not appear in the usual parton
densities, where $\zeta=0$.

In the case of matrix elements of space-like currents, one can choose
the special frame $(P-P')^+=0$, as in the Drell-Yan-West
representation of the space-like electromagnetic form factors
\cite{Drell:1970km}. Thus given the light-cone wavefunctions, one can
construct space-like electromagnetic, electroweak, gravitational
couplings, or any local operator product matrix element from their
overlap \cite{Brodsky:2000ii,Brodsky:1980zm}. This overlap is diagonal
in parton number. In the case of deeply virtual Compton scattering,
the proton matrix elements require the computation of the diagonal,
parton number conserving, matrix element $n \rightarrow n$ for the
regions $\zeta < x <1$ and $\zeta-1< x < 0$
\cite{Diehl:1999kh}. However, it also involves an off-diagonal
$n+1\rightarrow n-1$ convolution for $0< x < \zeta$, where the parton
number is decreased by two. This domain occurs since the current
operator of the final-state photon with positive light-cone momentum
fraction $\zeta$ can annihilate a quark-antiquark pair in the initial
proton wavefunction. This type of overlap has first been identified in
the context of the form factors which control time-like semi-leptonic
$B$ decay \cite{Brodsky:1999hn}.  As we shall see, there are
underlying relations between Fock states of different particle number
which interrelate the two types of overlap.

It also should be noted that the calculation of deep inelastic
structure functions and space-like form factors requires the
light-cone frame choice $(P-P')^+ \ne 0$ in one-space and one-time
theories for $(P-P')^2 \ne 0$.  Explicit non-perturbative results for
space-like form factors and structure functions of $QCD(1+1)$ with
$N_C \to \infty$ have been given by Einhorn \cite{Einhorn:1976uz}.
The application to deeply virtual Compton scattering in $QCD(1+1)$ has
recently been given by Burkardt \cite{Burkardt:2000uu}.

\begin{figure}[htb]
\begin{center}
\leavevmode
\epsfbox{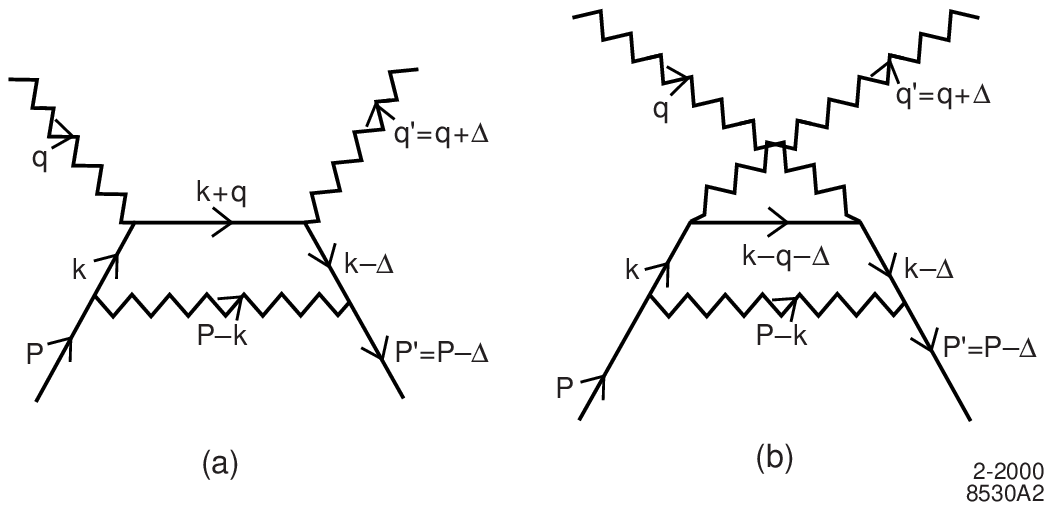}
\end{center}
\caption[*]{One-loop covariant Feynman diagrams for virtual Compton
scattering in QED.}
\label{fig:2}
\end{figure}

In order to illustrate the general formalism for $3+1$ theories, we
will present an explicit calculation of deeply virtual Compton
scattering on a fermion in quantum electrodynamics at one-loop
order. The Feynman amplitudes which are evaluated are shown in
Fig.~\ref{fig:2}.  The QED calculation
\cite{Brodsky:2000ii,Brodsky:1980zm} is patterned after the structure
which occurs in the one-loop Schwinger ${\alpha / 2 \pi} $ correction
to the electron magnetic moment.  In effect, we will represent a
spin-$\frac{1}{2}$ system as a composite of a spin-$\frac{1}{2}$
fermion and spin-one vector boson with arbitrary masses.  The one-loop
model illustrates the interrelations between Fock states of different
particle number as required by the boost invariance of space-like form
factors or, equivalently, by the $\zeta$ independence of the first
moment in $x$ of the generalized Compton form factors $H(x,\zeta,t)$
and $E(x,\zeta,t)$.

\vspace{.5cm}

The one-loop model can be further generalized by applying spectral
Pauli-Villars integration over the constituent masses. The resulting
form of light-cone wavefunctions provides a template for
parametrizing the structure of relativistic composite systems and
their matrix elements in hadronic physics. For example, this model has
recently been used to clarify the connection of parton distributions
to the constituents' spin and orbital angular momentum and to other
observables of the composite system such as its electromagnetic and
gravitational moments and form factors \cite{Brodsky:2000ii}. The
model also provides a self-consistent form for the wavefunctions of an
effective quark-diquark model of the valence Fock state of the proton
wavefunction.

This paper is organized as follows. After introducing the necessary
kinematics in Section~\ref{sec:kinematics}, we review in
Section~\ref{sec:form-factors} the representation of deeply virtual
Compton scattering in terms of generalized form factors, and the sum
rules connecting them with the form factors of the electromagnetic and
gravitational currents. In Section~\ref{sec:overlap} we represent the
generalized Compton form factors in terms of light-cone wavefunctions,
starting from the Fock state representation of a composite system. The
following section applies this framework to the explicit cases of
QED at one loop. We summarize our results in
Section~\ref{sec:con}. Throughout our paper we shall use momentum
variables as employed by Radyushkin \cite{Radyushkin:1997ki}, which
provide an intuitive parametrization in the context of wavefunction
overlaps. In the Appendix we give our main formulae in the momentum
variables of Ji \cite{Ji:1997ek}, which make the symmetry between the
incoming and outgoing proton more explicit.

\section{The Kinematics of Virtual Compton Scattering}
\label{sec:kinematics}

\vspace{.5cm}
\begin{figure}[htbp]
\begin{center}
\leavevmode
\epsfbox{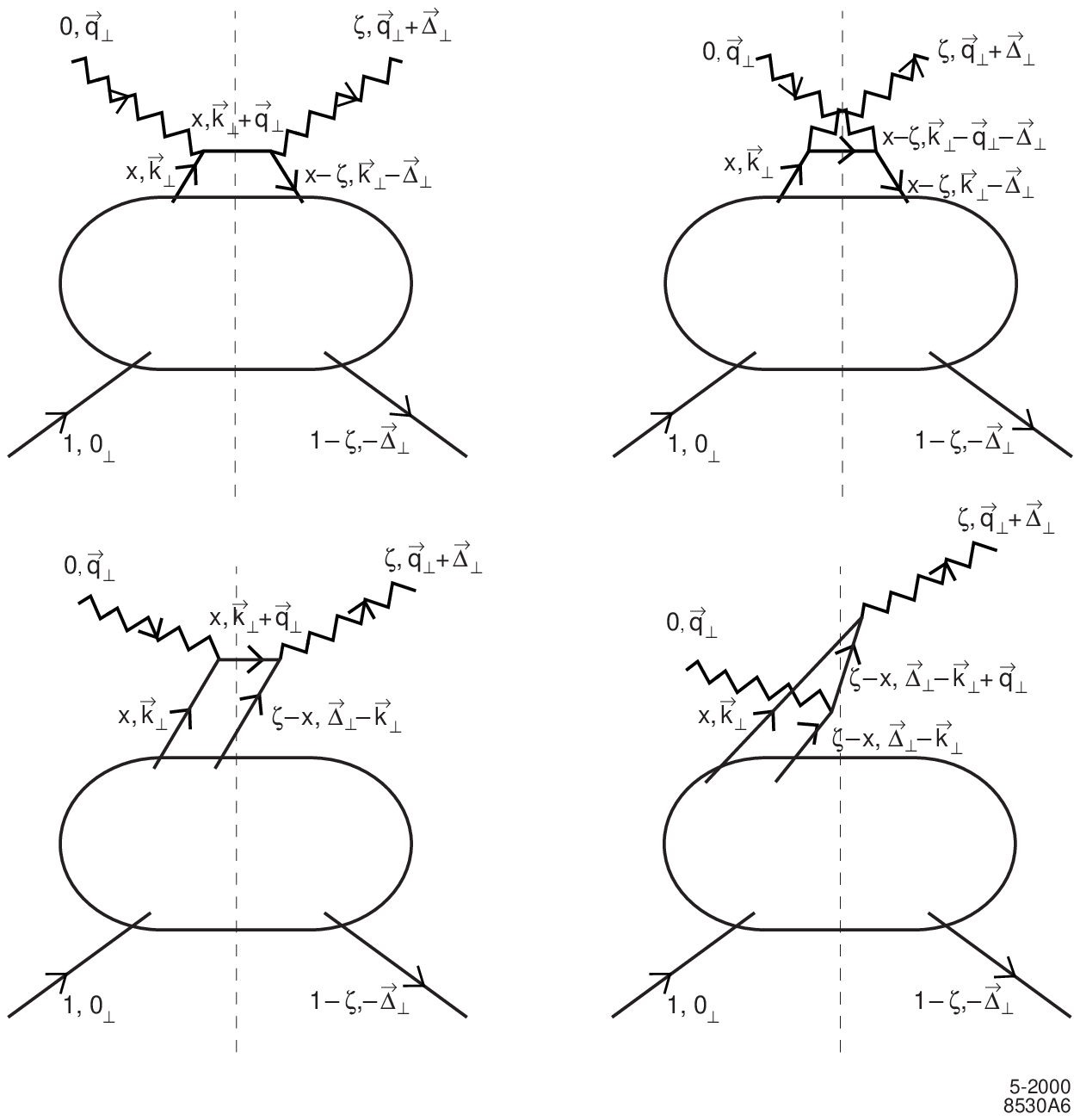}
\end{center}
\caption[*]{Light-cone time-ordered contributions to deeply virtual
Compton scattering. Only the contributions of leading power in $1/Q$
are illustrated.  These contributions illustrate the factorization
property of the leading twist amplitude.}
\label{fig:3}
\end{figure}

We begin with the kinematics of virtual Compton scattering
\begin{equation}
\gamma^*(q) + p(P) \to \gamma(q') + p(P')\ ,
\end{equation}
see Fig.~\ref{fig:3}. We specify the frame by choosing a convenient
parametrization of the light-cone coordinates for the initial and
final proton:
\begin{eqnarray}
P&=&
\left(\ P^+\ ,\ {\vec 0_\perp}\ ,\ {M^2\over P^+}\ \right)\ ,
\label{a1}\\
P'&=&
\left( (1-\zeta)P^+\ ,\ -{\vec \Delta_\perp}\ ,\ {M^2+{\vec
\Delta_\perp}^2 \over (1-\zeta)P^+}\right)\ ,
\end{eqnarray}
where $M$ is the proton mass. We use the component notation $V = (V^+,
\vec{V}_\perp, V^-)$, and our metric is specified by $V^\pm = V^0 \pm
V^z$ and $V^2 = V^+ V^- - {\vec V}_\perp^2$. The four-momentum
transfer from the target is
\begin{eqnarray}
\label{delta}
\Delta&=&P-P'\ =\
\left( \zeta P^+\ ,\ {\vec \Delta_\perp}\ ,\
{t+{\vec \Delta_\perp}^2 \over \zeta P^+}\right)\ ,
\end{eqnarray}
where $t = \Delta^2$. In addition, overall energy-momentum
conservation requires $\Delta^- = P^- - P'^-$, which connects ${\vec
\Delta_\perp}^2$, $\zeta$, and $t$ according to
\begin{equation}
 t \ = \ 2P\cdot \Delta\  =\
 -{\zeta^2M^2+{\vec \Delta_\perp}^2 \over 1-\zeta}\ .
 \label{na1anew}
\end{equation}

As in the case of space-like form factors, it is convenient to choose
a frame where the incident space-like photon carries $q^+ = 0$ so that
$q^2= - Q^2 = - {\vec q_\perp}^{\;2}$:
\begin{eqnarray}
q&=& \left( 0\ ,\ {\vec q_\perp}\ ,\
{({\vec q_\perp}+{\vec \Delta_\perp})^2\over \zeta P^+}
+{\zeta M^2+{\vec \Delta_\perp}^2 \over (1-\zeta)P^+}\right)\ ,
\label{a2}
\\
q'&=&
\left( \zeta P^+\ ,\ {\vec q_\perp}+{\vec \Delta_\perp}\ ,\
{({\vec q_\perp}+{\vec \Delta_\perp})^2\over \zeta P^+}\right)\ .
\label{a2p}
\end{eqnarray}
Thus no light-cone time-ordered amplitudes involving the splitting of
the incident photon can occur. The variable $\zeta$ is fixed from
(\ref{a1}) and (\ref{a2}) as
\begin{equation}
2P\cdot q={({\vec q_\perp}+{\vec \Delta_\perp})^2\over \zeta }
+{\zeta M^2+{\vec \Delta_\perp}^2 \over 1-\zeta}\ .
\label{nn2}
\end{equation}
We will be interested in deeply virtual Compton scattering, where
$Q^2$ is large compared to the masses and $-t$. Then, we have
\begin{equation}
{Q^2\over 2P \cdot q}=\zeta\
\label{nn3}
\end{equation}
up to corrections in $1/Q^2$. Thus $\zeta$ plays the role of the
Bjorken variable in deeply virtual Compton scattering. For a fixed
value of $-t$, the allowed range of $\zeta$ is given by
\begin{equation}
0\ \le\ \zeta\ \le\
{(-t)\over 2M^2}\ \ \left( {\sqrt{1+{4M^2\over (-t)}}}\ -\ 1\ \right)\ .
\label{nn4}
\end{equation}
The choice of parametrization of the light-cone frame is of course
arbitrary.  For example, in the Appendix, we show how one can
conveniently utilize a ``symmetric'' frame for the incoming and
outgoing proton, which has manifest symmetry under crossing $P
\leftrightarrow P'$.

\section{The Generalized Form Factors of Deeply Virtual Compton
Scattering}
\label{sec:form-factors}

The virtual Compton amplitude $M^{\mu\nu}({\vec q_\perp},{\vec
\Delta_\perp},\zeta)$, {\it i.e.}, the transition matrix element of
the process $\gamma^*(q) + p(P) \to \gamma(q') + p(P')$, can be
defined from the light-cone time-ordered product of currents
\begin{eqnarray}
&&M^{\mu\nu}({\vec q_\perp},{\vec \Delta_\perp},\zeta)\ =\
i\int d^4y\,
e^{-iq\cdot y}\langle P'|TJ^\mu (y)J^\nu (0)|P\rangle \ ,
\label{com1j}
\end{eqnarray}
where the Lorentz indices $\mu$ and $\nu$ denote the polarizations of
the initial and final photons respectively.  An essential
characteristic of deeply virtual Compton scattering in light-cone
gauge is that any soft interaction with the target which occurs
between the light-cone times of the incident and final photon is
power-law suppressed as $1/Q$. Similarly, the diagrams in which the
photons hit two different quark lines in the target are also higher
twist.  We can then replace the fully interacting currents $J^\mu (y)$
and $J^\nu (0)$ by the quark currents $j^\mu (y)$ and $j^\nu (0)$ of
the non-interacting theory, which have simple matrix elements in the
free Fock basis.  The leading contribution thus factorizes as a
hard-scattering amplitude involving the elementary photon interactions
on a quark line convoluted with the non-perturbative light-cone
wavefunctions of the protons, see Fig.~\ref{fig:3}. In the limit
$Q^2\to \infty$ at fixed $\zeta$ and $t$ the Compton amplitude is thus
given by
\begin{eqnarray}
\lefteqn{
M^{IJ}({\vec q_\perp},{\vec \Delta_\perp},\zeta)\ =\
\epsilon^I_\mu\, \epsilon^{*J}_\nu\,
M^{\mu\nu}({\vec q_\perp},{\vec \Delta_\perp},\zeta)\ =\
- e^2_{q}\ {1 \over 2\bar P^+}
\int_{\zeta-1}^1{\rm d}x\
}
\label{com1a}
\\
&\times& \left\{ \ {t}^{IJ}(x,\zeta)\ {\bar U}(P')
\left[
H(x,\zeta,t)\ {\gamma^+}
 +
E(x,\zeta,t)\
{i\over 2M}\, {\sigma^{+\alpha}}(-\Delta_\alpha)
\right] U(P) \right.
\nonumber\\
&& \ \
\left. {s}^{IJ}(x,\zeta)\ {\bar U}(P')
\left[
{\widetilde H}(x,\zeta,t)\ {\gamma^+\gamma_5}
 +
{\widetilde E}(x,\zeta,t)\
{1\over 2M}\, {\gamma_5}(-\Delta^+)
\right] U(P)\
\right\} ,
\nonumber
\end{eqnarray}
where $\bar P={1\over 2}(P'+P)$. In getting (\ref{com1a}) we used the
relations $\gamma^1\gamma^+\gamma^1=\gamma^2\gamma^+\gamma^2=\gamma^+$
and
$\gamma^1\gamma^+\gamma^2=-\gamma^2\gamma^+\gamma^1=i\gamma^+\gamma_5$.
For simplicity we only consider one quark with flavor $q$ and electric
charge $e_{q}$ here and in the sequel. Throughout our analysis we will
use the Born approximation to the photon-quark amplitude. The
radiative QCD corrections to this process have been calculated to
order $\alpha_s$ in \cite{Ji:1998xh,Mankiewicz:1998bk}.

For circularly polarized initial and final photons ($I,\ J$ are
$\uparrow$ or $\downarrow)$) we have
\begin{eqnarray}
{t}^{\ \uparrow\uparrow}(x,\zeta)&=&
\phantom{-} \ {t}^{\ \downarrow\downarrow}(x,\zeta)\ =\
{1\over x-i\epsilon}\
+\ {1\over x-\zeta +i\epsilon}\ ,
\label{com2p}\\
{s}^{\ \uparrow\uparrow}(x,\zeta)&=&
-\ {s}^{\ \downarrow\downarrow}(x,\zeta)\ =\
{1\over x-i\epsilon}\ -\ {1\over x-\zeta +i\epsilon}\ ,
\nonumber\\
{t}^{\ \uparrow\downarrow}(x,\zeta)
&=& {t}^{\ \downarrow\uparrow}(x,\zeta)
\ =\ {s}^{\ \uparrow\downarrow}(x,\zeta)
\ =\ {s}^{\ \downarrow\uparrow}(x,\zeta)
\ \ =\ \ 0\ \phantom{\frac{1}{2}}.
\nonumber
\end{eqnarray}
The two photon polarization vectors in light-cone gauge are given by
\begin{equation}
\epsilon^{\uparrow,\downarrow}=
\left(0\ ,\ \vec \epsilon^{\ \uparrow,\downarrow}_\perp\ ,\
{\vec \epsilon^{\ \uparrow,\downarrow}_\perp
 \cdot \vec k_\perp \over 2 k^+} \right)\ ,
\qquad
\vec \epsilon_\perp^{\ \uparrow,\downarrow}=
\mp {1\over\sqrt{2}} \left(
\begin{array}{c}
1 \\ \pm i
\end{array}
\right) \ ,
\label{p-vectors}
\end{equation}
where $k$ denotes the appropriate photon momentum. The polarization
vectors satisfy the Lorentz condition $ k \cdot \epsilon =0$. For a
longitudinally polarized initial photon, the Compton amplitude is of
order $1/Q$ and thus vanishes in the limit $Q^2\to \infty$. At order
$1/Q$ there are several corrections to the simple structure in
Eq.~(\ref{com1a}). Those corrections which correspond to the
evaluation of the diagrams in Fig.~\ref{fig:3} to accuracy $1/Q$ can
be described by functions related to the form factors $H$, $E$ or
${\widetilde H}$, ${\widetilde E}$ through Wandzura-Wilczek type
integral relations. In other contributions the hard scattering is no
longer on a single quark line, so that new non-perturbative functions
appear. For details we refer to~\cite{Blumlein:1999sc,Anikin:2000em}.

In (\ref{com1a}) the generalized form factors $H$, $E$ and
${\widetilde H}$, ${\widetilde E}$ are defined through matrix elements
of the bilinear vector and axial vector currents on the light-cone:
\begin{eqnarray}
\lefteqn{
\int\frac{d y^-}{8\pi}\;e^{ix P^+y^-/2}\;
\langle P' | \bar\psi(0)\,\gamma^+\,\psi(y)\,|P\rangle
\Big|_{y^+=0, y_\perp=0}
} \hspace{2em}
\label{spd-def} \\
&=&
{1\over 2\bar P^+}\ {\bar U}(P') \left[ \,
H(x,\zeta,t)\ {\gamma^+}
 +
E(x,\zeta,t)\
{i\over 2M}\, {\sigma^{+\alpha}}(-\Delta_\alpha)
\right]  U(P)\ ,
\nonumber\\
\lefteqn{
\int\frac{d y^-}{8\pi}\;e^{ix P^+y^-/2}\;
\langle P' | \bar\psi(0)\,\gamma^+\gamma_5\,\psi(y)\,|P\rangle
\Big|_{y^+=0, y_\perp=0}
} \hspace{2em}
\nonumber\\
&=&
{1\over 2\bar P^+}\ {\bar U}(P') \left[ \,
{\widetilde H}(x,\zeta,t)\ {\gamma^+\gamma_5}
 +
{\widetilde E}(x,\zeta,t)\
{1\over 2M}\, {\gamma_5}(-\Delta^+)\,
\right]  U(P)\ .
\nonumber
\end{eqnarray}

We can compare these generalized form factors for the Compton
amplitude with the matrix elements of the electromagnetic and
gravitational currents. For the electromagnetic current
$e_q J^\mu(y)=e_{q}{\bar{\psi}}(y)\gamma^\mu \psi(y)$, one has the
usual Dirac and Pauli form factors
\begin{equation}
\langle P'| J^\mu(0) |P\rangle
= \bar U(P') \left[\, F_{1}(t)\,  \gamma^\mu +
F_{2}(t)\, {i\over 2M}\, \sigma^{\mu\alpha}(-\Delta_\alpha)\,
\right] U(P)\ ,
\label{Drell1}
\end{equation}
where for later convenience we have not included the charge $e_q$ in
the definitions of the current and the form factors. For the form
factors of the energy-momentum tensor for a spin-${1\over 2}$
composite system, one defines~\cite{Ji:1997ek,Kobzarev:1963}
\begin{eqnarray}
\langle P'| T^{\mu\nu} (0)|P \rangle
&=& \bar U(P') \left[\, A_{q}(t)\,
{\gamma}^{(\mu}\,  \bar P^{\nu)} +
B_{q}(t)\, {i\over 2M} \bar P^{(\mu} {\sigma}^{\nu)\alpha}
(-\Delta_\alpha) \right.\nonumber \\
&&\qquad\qquad + \left. C_{q}(t)\,
{1\over M}\, (\Delta^\mu \Delta^\nu - g^{\mu\nu}\Delta^2)
\, \right] U(P) \ ,
\label{Ji12}
\end{eqnarray}
where $a^{(\mu}b^{\nu)}={1\over 2}(a^\mu b^\nu +a^\nu b^\mu)$ and
\begin{equation}
T^{\mu\nu} = {i\over 4}\ \Big(\
 {\bar{\psi}}\gamma^\mu \overrightarrow{\partial}{}^\nu \psi
-{\bar{\psi}}\gamma^\mu \overleftarrow{\partial}{}^\nu \psi
\ \Big)
\ +\ \{\ \mu\longleftrightarrow\nu\ \}\
\label{lagsva}
\end{equation}
is the quark part of the energy-momentum tensor. These form factors
can be computed in a form similar to that of the virtual Compton
amplitude if we choose the light-cone frame where the virtual photon
or graviton has momentum $q^+ = \zeta P^+$ and ${\vec q_\perp}={\vec
\Delta_\perp}$. In the electromagnetic case, the coupling of the
current $e_q J^+(0)$ on the quark line is identical to the Compton
amplitude with $e_{q}^2\, t^{IJ}$ replaced simply by the quark charge
$e_{q}$. One finds
\begin{eqnarray}
\int_{\zeta-1}^1 {{\rm d}x\over 1-\zetahalf}\ H(x,\zeta,t)&=&
 F_{1}(t) \ ,
\label{com3}\\
\int_{\zeta-1}^1 {{\rm d}x\over 1-\zetahalf}\ E(x,\zeta,t)&=&
 F_{2}(t) \ ,
\nonumber
\end{eqnarray}
Analogous sum rules relate $\widetilde H$ and $\widetilde E$ with the
form factors of the axial vector current
$J_5^\mu(y)={\bar{\psi}}(y)\gamma^\mu\gamma_5 \psi (y)$. The factors
$1-\zeta/2$ in (\ref{com3}) appear because we use Ji's normalization
convention for the Compton form factors, which involves $\bar P^+$ on
the right-hand side of (\ref{spd-def}), and at the same time
parametrize light-cone momentum fractions with respect to
$P^+=(1-\zeta/2) \bar P^+$. In the Appendix we will give our main
formulae in the parametrization of Ji, where momentum fractions refer
to $\bar P^+$.

In the case of the gravitational form factors, the derivative coupling
in the graviton current $T^{++}$ brings in an extra factor $x -
\zeta/2$. Then, one gets the sum rule \cite{Ji:1997ek}
\begin{equation}
\int_{\zeta-1}^1 {{\rm d}x\over 1-\zetahalf}\
{x - \zetahalf \over 1-\zetahalf}\
[H(x,\zeta,t) + E(x,\zeta,t)] \ =\
A_{q}(t) \ +\ B_{q}(t)\ .
\label{com3zaa}
\end{equation}
The gravitational form factor $C_{q}(t)$ cancels for the combination
$H+E$, as we shall show shortly.

When we take the $\mu =+$ component in (\ref{Drell1}), the factors of
the two terms in the right hand side are given by
\begin{eqnarray}
{1 \over 2\bar P^+}\
{\bar U}(P',{\lambda}'){\gamma^+}\, U(P,\lambda) &=&
{\sqrt{1-\zeta} \over 1-\zetahalf }\ \delta_{\lambda ,\, {\lambda}'}\ ,
\label{com1z}\\
{1 \over 2\bar P^+}\
{\bar U}(P',{\lambda}'){i \over 2M}\,
{\sigma^{+\alpha}}(-\Delta_\alpha) U(P,\lambda)
&=&
-\, {\zeta^2 \over 4 {(1-\zetahalf) \sqrt{1-\zeta}}}\
    \delta_{\lambda ,\, {\lambda}'}
\nonumber\\
&& +\
{1 \over \sqrt{1-\zeta}} \;
{\, -\lambda \Delta^1\, -\, i\Delta^2\, \over 2M}\
\delta_{\lambda ,\, -{\lambda}'}\ ,
\nonumber
\end{eqnarray}
where $\lambda=\pm 1$ is the light-cone helicity of the initial
fermion. Then, in the light-cone formalism, the Dirac and Pauli form
factors can be identified from the helicity-conserving and
helicity-flip vector current matrix elements:
\begin{eqnarray}
\VEV{P',\uparrow\left|\frac{J^+(0)}{2\bar P^+}
\right|P,\uparrow} &=&
{\sqrt{1-\zeta} \over 1-\zetahalf}\, F_{1}(t)\,
-\, {\zeta^2 \over 4 (1-\zetahalf) {\sqrt{1-\zeta}}}\, F_{2}(t)\ ,
\label{BD1}\\
\VEV{P',\uparrow\left|\frac{J^+(0)}{2\bar P^+}\right|P,\downarrow}
&=&
{1 \over \sqrt{1-\zeta}} \;
{\Delta^1-{i} \Delta^2\over 2M}\, F_{2}(t)\ .
\label{BD2}
\end{eqnarray}
For the matrix element (\ref{Ji12}) of the energy-momentum tensor we
also need
\begin{equation}
{1 \over 2M}\ {\bar U}(P',{\lambda}')U(P,\lambda)\ =\
{1-\zetahalf \over {\sqrt{1-\zeta}}}\
\delta_{\lambda ,\, {\lambda}'}
-{1\over {\sqrt{1-\zeta}}} \;
{-\lambda \Delta^1\, -\, i\Delta^2 \over 2M}\
\delta_{\lambda ,\, -{\lambda}'}\ .
\label{ab1}
\end{equation}
One easily checks that (\ref{com1z}) and (\ref{ab1}) satisfy the
Gordon identity. With this we have
\begin{eqnarray}
\VEV{P',\uparrow\left|\frac{T^{++}(0)}{2(\bar P^+)^2}
\right|P,\uparrow} &=&
{\sqrt{1-\zeta} \over 1-\zetahalf}\, A_{q}(t)\,
-\, {\zeta^2 \over 4 (1-\zetahalf) {\sqrt{1-\zeta}}}\, B_{q}(t)
\nonumber\\
&&+\ {\zeta^2\over (1-\zetahalf) {\sqrt{1-\zeta}}}\, C_{q}(t)\ ,
\label{BD1t}\\
\VEV{P',\uparrow\left|
   \frac{T^{++}(0)}{2(\bar P^+)^2}\right|P,\downarrow}
&=&
\left\{ {1 \over {\sqrt{1-\zeta}}}\,
B_{q}(t) \right.
\nonumber\\
&& \left.
-\ {\zeta^2\over (1-\zetahalf)^2 {\sqrt{1-\zeta}}}\,
C_{q}(t)\ \right\}
\ {\Delta^1-{i} \Delta^2\over 2M}\ .
\label{BD2t}
\end{eqnarray}
Combining (\ref{BD1t}) and (\ref{BD2t}), we get
\begin{eqnarray}
&&{1\over 1 - \zetahalf}
\VEV{P',\uparrow\left|
   \frac{T^{++}(0)}{2(\bar P^+)^2}\right|P,\uparrow}
+ {2M \over  {\Delta^1-{i} \Delta^2}}
\VEV{P',\uparrow\left|
   \frac{T^{++}(0)}{2(\bar P^+)^2}\right|P,\downarrow}
\nonumber\\
&&=
{\sqrt{1-\zeta} \over (1-\zetahalf)^2}\ \left[ A_{q}(t)\ + \
B_{q}(t) \right]\ ,
\label{a706a1}
\end{eqnarray}
whereas from (\ref{BD1}) and (\ref{BD2}) we have
\begin{eqnarray}
&&{1\over 1 - \zetahalf}
\VEV{P',\uparrow\left|
   \frac{J^{+}(0)}{2\bar P^+}\right|P,\uparrow}
+ {2M \over  {\Delta^1-{i} \Delta^2}}
\VEV{P',\uparrow\left|
   \frac{J+(0)}{2\bar P^+}\right|P,\downarrow}
\nonumber\\
&&=\
{\sqrt{1-\zeta} \over (1-\zetahalf)^2}\ \left[ F_{1}(t)\ + \
F_{2}(t) \right]\ .
\label{a706a2}
\end{eqnarray}
Using that $H$ and $E$ involve the same proton helicity structure as
$F_{1}$ and $F_{2}$, see (\ref{spd-def}) and (\ref{Drell1}), one
obtains Ji's sum rule (\ref{com3zaa}) from the connection between
$T^{++}(0)$ and the non-local current defining $H$ and $E$.

\begin{figure}[htb]
\begin{center}
\leavevmode
\epsfbox{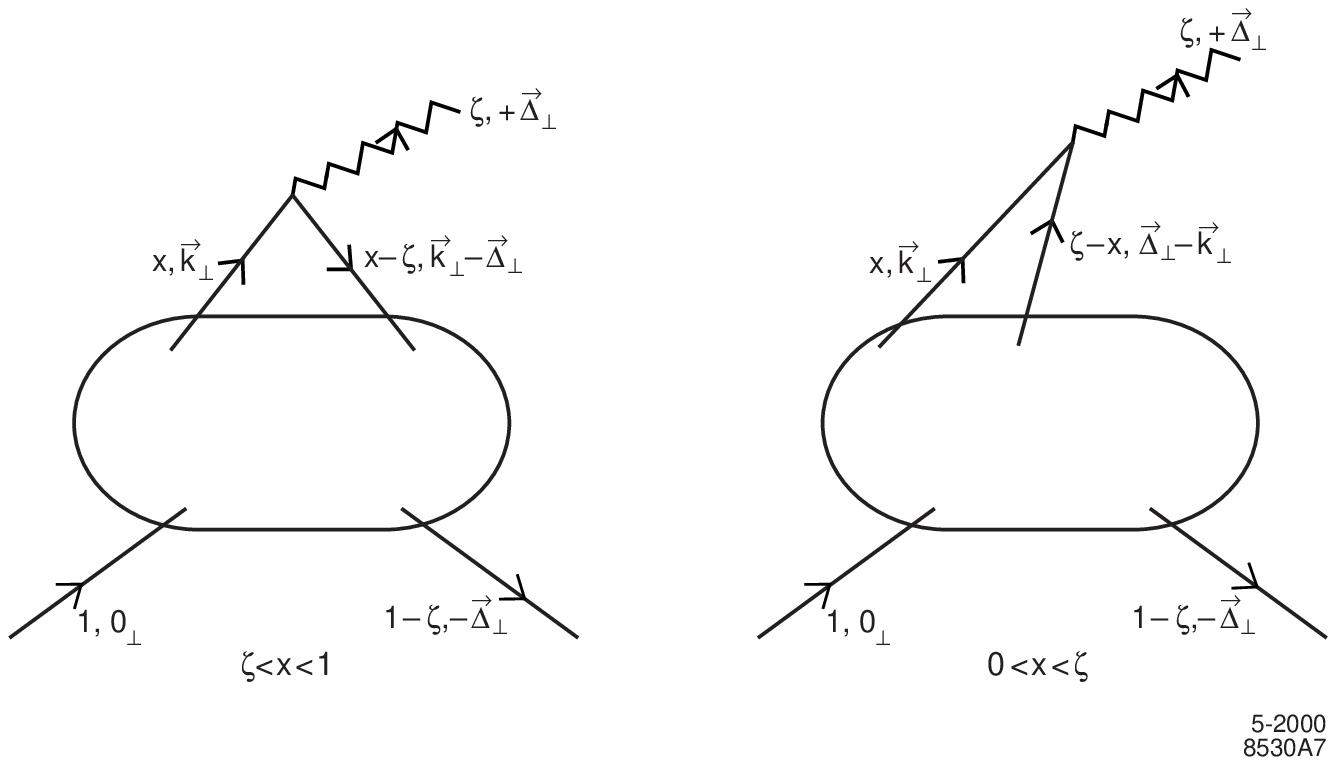}
\end{center}
\caption[*]{Light-cone time-ordered contributions to spacelike form
factors.  The sum of the two contributions is $\zeta$-independent at
fixed $t = \Delta^2$.}
\label{fig:4}
\end{figure}

In (\ref{com3}) we can separate three distinct regions of integration:
\begin{eqnarray}
\lefteqn{
(1-\zeta/2)\, F_{1}(t) \rule[-2ex]{0pt}{2ex}
}
\label{d1} \\
&=&\int_{\zeta-1}^0{\rm d}x\ H_{(n\to n)}(x,\zeta,t)
+\int_{0}^\zeta {\rm d}x\ H_{(n+1 \to n-1)}(x,\zeta,t)
+\int_{\zeta}^1{\rm d}x\ H_{(n\to n)}(x,\zeta,t) \,
\nonumber
\end{eqnarray}
where
\begin{eqnarray}
H(x,\zeta,t) &=&
H_{(n\to n)}(x,\zeta,t)\ [ \theta(x-\zeta) + \theta(-x) ]
\\
&+& H_{(n+1 \to n-1)}(x,\zeta,t)\ \theta(\zeta-x)\, \theta(x)\ .
\nonumber
\end{eqnarray}
Similarly,
\begin{eqnarray}
\lefteqn{
(1-\zeta/2)\, F_{2}(t) \rule[-2ex]{0pt}{2ex}
}
\label{d3} \\
&=&\int_{\zeta-1}^0{\rm d}x\ E_{(n\to n)}(x,\zeta,t)
+\int_{0}^\zeta {\rm d}x\ E_{(n+1 \to n-1)}(x,\zeta,t)
+\int_{\zeta}^1{\rm d}x\ E_{(n\to n)}(x,\zeta,t) \,
\nonumber
\end{eqnarray}
with
\begin{eqnarray}
E(x,\zeta,t) &=&
 E_{(n\to n)}(x,\zeta,t)\ [ \theta(x-\zeta) + \theta(-x) ]
\\
&+& E_{(n+1 \to n-1)}(x,\zeta,t)\ \theta(\zeta-x)\, \theta(x)\ .
\nonumber
\end{eqnarray}
{}From the point of view of the form factors $F_{1}(t)$ and
$F_{2}(t)$, the division of the $x$-integrals into separate domains
$\zeta-1<x<0$, $0< x < \zeta$, and $\zeta < x < 1$ is an artifact of
the light-cone frame choice. The two types of contributions to
space-like form factors are illustrated in Fig.~\ref{fig:4}. In the
region $\zeta< x < 1$ a quark is scattered off the current, whereas
the region $\zeta-1<x<0$ corresponds to scattering of an
antiquark. For $0<x<\zeta$, however, the current annihilates a
quark-antiquark pair in the target.

The $x$-integrals of $(1-\zeta/2)^{-1}\, H(x,\zeta,t)$ and
$(1-\zeta/2)^{-1}\, E(x,\zeta,t)$ are independent of $\zeta$ as a
consequence of Lorentz invariance. Thus, the contributions of the $n$,
$n-1$, and $n+1$ particle light-cone wavefunctions are in fact not
independent. This deeply reflects the underlying frame invariance of
the light-cone Fock representation \cite{Glazek:1990rr}, which ensures
the independence of the form factors from the frame choice for the
graviton or photon momentum $\Delta^+ = \zeta P^+$ as long as
$\Delta^2 = t$ is kept fixed.

\begin{figure}[htbp]
\begin{center}
\leavevmode
\epsfbox{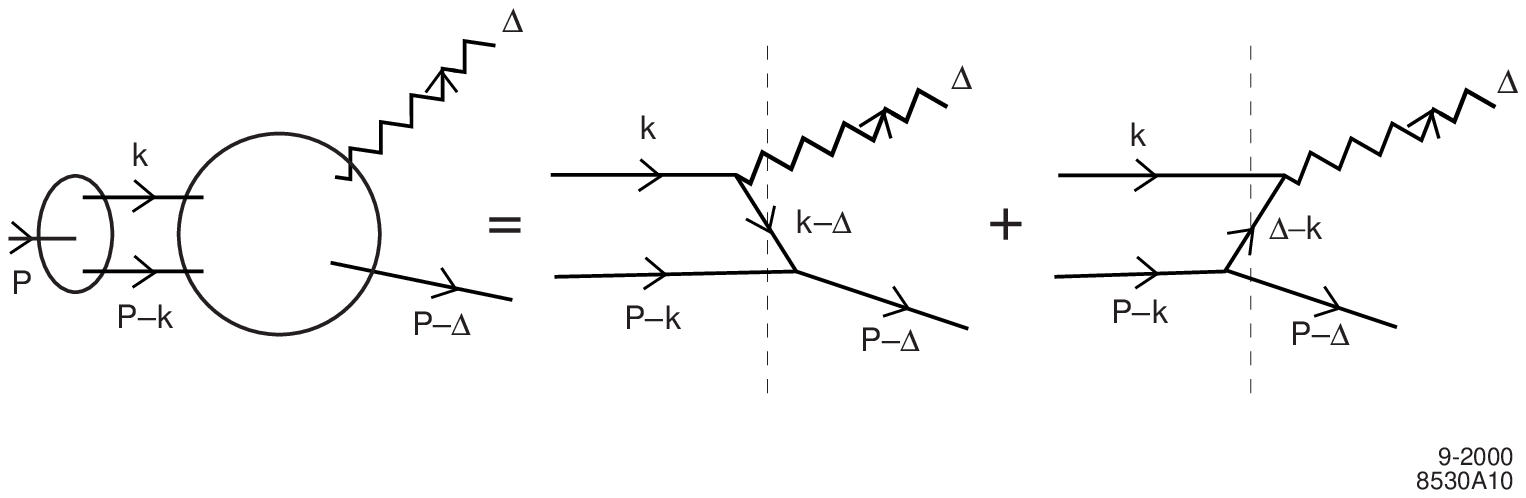}
\end{center}
\caption[*]{Light-cone time-ordered contributions from $t$-channel
exchange in the scattering process $k + (P-k) \to \Delta + (P-\Delta)$
in $\phi^3$ theory.}
\label{fig:4a}
\end{figure}

To see how this occurs in a simple example, we shall consider the
matrix element for the scattering of two scalar particles, $k + (P-k)
\to \Delta + (P-\Delta)$, in $\phi^3$ theory. In order to make the
result resemble the integrand of a form factor, we parametrize $k^+=
xP^+$ and $\Delta^+=\zeta P^+$. The particle with momentum $\Delta$
represents the external current, whereas the other external particles
are on their mass shell, $k^2 = (P-k)^2 = (P-\Delta)^2 =
m^2$. Four-momentum conservation implies $k^- + (P-k)^- = \Delta^- +
(P-\Delta)^-$. The Born amplitude from exchanging a scalar particle
with momentum $\Delta-k$ in the $t$-channel receives two time-ordered
contributions in light-cone perturbation theory, see
Fig.~\ref{fig:4a}:
\begin{eqnarray}
{\cal M}_{k + (P-k) \to \Delta + (P-\Delta)} &=&
{g^2 \theta(\zeta - x) \over
(\Delta-k)^+ \left[ (P-k)^- - (P-\Delta)^- -
     {(\vec\Delta_\perp-\vec k_\perp)^2 + m^2 \over (\Delta-k)^+} + i
\epsilon\right]} \ \ \
\\
&+& {g^2 \theta(x - \zeta)
\over (k-\Delta)^+ \left[ k^- - \Delta^- -
     {(\vec k_\perp-\vec\Delta_\perp)^2 + m^2 \over (k-\Delta)^+} + i
\epsilon\right]}\ .
\nonumber
\end{eqnarray}
Since the particle with momentum $\Delta$ represents the external
current we have treated it as an on-shell particle with mass squared
$\Delta^2$, so that $\Delta^- = (\vec\Delta_\perp^2 + \Delta^2)
/\Delta^+$ appears in the energy denominator of the second term
\cite{Lepage:1980fj}. Then, the sum of the light-cone contributions
must agree with the covariant result
\begin{equation}
{\cal M}_{k + (P-k) \to \Delta + (P-\Delta)} \ = \
 {g^2 \over (k-\Delta)^2 -m^2 + i \epsilon} \,
\end{equation}
independent of $\zeta$. In fact, replacing $(P-k)^- - (P-\Delta)^-$
with $\Delta^- - k^-$ in the first term, we see that the identity is
trivial since the two denominators in ${\cal M}$ are identical,
differing only in whether $0<x<\zeta$ or $\zeta<x<1$ as determined by
the arbitrary choice of frame.

The scattering amplitude can also be interpreted as a simple model for
a scalar current form factor $\VEV{P-\Delta\vert J(0)\vert P}.$ In
this model, one interprets the initial state $| P\rangle$ as a wave
packet of the incident particles with momenta $P-k$ and $k$.  The
particle with momentum $\Delta$ represents again the particle coupling
to the current. The two light-cone time-ordered contributions of $\cal
M$ correspond to the $3 \to 1$ and $2 \to 2$ light-cone wavefunction
overlap contributions to the form factor, respectively.  Thus in this
simple model, the division of contributions from the $3 \to 1$ and $2
\to 2$ light-cone wavefunctions corresponds simply to the partition of
a common $x$ integrand into regions $0<x<\zeta$ and $\zeta<x<1$.

\section{The Light-Cone Fock Representation of Deeply Virtual Compton
Scattering}
\label{sec:overlap}

The light-cone Fock expansion of hadrons is constructed by quantizing
QCD at fixed light-cone time $\tau = t + z/c$ and forming the
invariant light-cone Hamiltonian: $ H_{LC} = P^+ P^- - {\vec
P_\perp}^2$, see \cite{Dirac:1949cp}.  While the momentum generators
$P^+$ and $\vec P_\perp$ are kinematical, {\it i.e.}, they are
independent of the interactions, the generator $P^- = i {{\mathrm
d}\over {\mathrm d}\tau}$ generates light-cone time translation. The
eigen-spectrum of $H_{LC}$ gives the entire mass spectrum of the
color-singlet hadron states in QCD, together with their respective
light-cone wavefunctions.  In particular, the proton state satisfies
$H_{LC} \ket{\psi_p} = M^2 \ket{\psi_p}$.  Such equations can be
solved in principle using the discretized light-cone quantization
(DLCQ) method \cite{Pauli:1985ps}. The expansion of the proton
eigensolution $\ket{\psi_p}$ on the eigenstates $\{\ket{n} \}$ of the
free Hamiltonian $ H_{LC}(g = 0)$ gives the light-cone Fock expansion:
\begin{eqnarray}
\left\vert \psi_p(P^+, {\vec P_\perp} )\right> &=& \sum_{n}\
\prod_{i=1}^{n}
  {{\rm d}x_i\, {\rm d}^2 {\vec k_{\perp i}}
\over \sqrt{x_i}\, 16\pi^3}\ \,
16\pi^3 \delta\left(1-\sum_{i=1}^{n} x_i\right)\,
\delta^{(2)}\left(\sum_{i=1}^{n} {\vec k_{\perp i}}\right)
\label{a318}
\\
&& \qquad \rule{0pt}{4.5ex}
\times \psi_n(x_i,{\vec k_{\perp i}},
\lambda_i) \left\vert n;\,
x_i P^+, x_i {\vec P_\perp} + {\vec k_{\perp i}}, \lambda_i\right>.
\nonumber
\end{eqnarray}
The light-cone momentum fractions $x_i = k^+_i/P^+$ and ${\vec
k_{\perp i}}$ represent the relative momentum coordinates of the QCD
constituents. The physical transverse momenta are ${\vec p_{\perp i}}
= x_i {\vec P_\perp} + {\vec k_{\perp i}}.$ The $\lambda_i$ label the
light-cone spin projections $S^z$ of the quarks and gluons along the
quantization direction $z$. The $n$-particle states are normalized as
\begin{equation}
\left< n;\, p'_i{}^+, {\vec p\,'_{\perp i}}, \lambda'_i \right. \,
\left\vert n;\,
p^{~}_i{}^{\!\!+}, {\vec p^{~}_{\perp i}}, \lambda_i\right>
= \prod_{i=1}^n 16\pi^3
  p_i^+ \delta(p'_i{}^{+} - p^{~}_i{}^{\!\!+})\
  \delta^{(2)}( {\vec p\,'_{\perp i}} - {\vec p^{~}_{\perp i}})\
  \delta_{\lambda'_i \lambda^{~}_i}\ .
\label{normalize}
\end{equation}
Here and in the following we will not display the other quantum
numbers of the partons, \textit{i.e.}, color and quark flavor. We will
also not discuss the case where a Fock state contains partons with
identical helicity, flavor, and color. For a discussion of these
points we refer to \cite{Diehl:2000xz}.

The solutions of $ H_{LC} \ket{\psi_p} = M^2 \ket{\psi_p}$ are
independent of $P^+$ and ${\vec P_\perp}$; thus given the
eigensolution Fock projections $ \langle n;\ x_i, {\vec k_{\perp i}},
\lambda_i |\psi_p \rangle \propto \psi_n(x_i, {\vec k_{\perp i}},
\lambda_i)$, the wavefunction of the proton is determined in any frame
\cite{Lepage:1980fj}. The light-cone wavefunctions $\psi_{n}(x_i,\vec
k_{\perp i},\lambda_i)$ encode all of the bound state quark and gluon
properties of hadrons, including their momentum, spin and flavor
correlations, in the form of universal process- and frame-independent
amplitudes.

The deeply virtual Compton amplitude can be evaluated explicitly by
starting from the Fock state representation for both the incoming and
outgoing proton, using the boost properties of the light-cone
wavefunctions, and evaluating the matrix elements of the currents for
a quark target. One can also directly evaluate the non-local current
matrix elements (\ref{spd-def}) in the same framework. In the
following we will concentrate on the generalized Compton form factors
$H$ and $E$. Formulae analogous to our results can be obtained for
$\widetilde{H}$ and $\widetilde{E}$.

For the $n \to n$ diagonal term ($\Delta n = 0$), the relevant current
matrix element at quark level is
\begin{eqnarray}
\lefteqn{
\int\frac{d y^-}{8\pi}\;e^{ix P^+y^-/2}\;
\left< 1;\, x'_{1} P\,'^+, {\vec p\,'_{\perp 1}, \lambda'_{1}}
  \left| \bar\psi(0)\, {\gamma^+}\,\psi(y)\, \right|
1;\, x^{~}_{1} P^+, {\vec p^{~}_{\perp 1}}, \lambda^{~}_{1}\right>
\Big|_{y^+=0, y_\perp=0}
} \hspace{14em}
\\ \nonumber
&=& \sqrt{x^{~}_{1} x'_{1}}\,\sqrt{1-\zeta}\ \delta(x-x_{1})\
    \delta_{\lambda'_{1} \lambda^{~}_{1}}\ ,
\end{eqnarray}
where for definiteness we have labeled the struck quark with the index
$i=1$. We thus obtain formulae for the diagonal
(parton-number-conserving) contributions to $H$ and $E$ in the domain
$\zeta\le x\le 1$ \cite{Diehl:1999kh}:
\begin{eqnarray}
\lefteqn{
{\sqrt{1-\zeta} \over 1-\zetahalf}\, H_{(n\to n)}(x,\zeta,t)\,
-\, {\zeta^2 \over 4 (1-\zetahalf) {\sqrt{1-\zeta}}} \,
    E_{(n\to n)}(x,\zeta,t)
} \hspace{5em}
\label{t1} \\
&=&
{\sqrt{1-\zeta}^{\, 2-n}}\, \sum_{n, \lambda_i}
\int \prod_{i=1}^{n}
{{\rm d}x_{i}\, {\rm d}^2{\vec{k}}_{\perp i} \over 16\pi^3 }\ \,
16\pi^3 \delta\left(1-\sum_{j=1}^n x_j\right) \, \delta^{(2)}
\left(\sum_{j=1}^n {\vec{k}}_{\perp j}\right)
\nonumber\\
&& \qquad \qquad \rule{0pt}{3ex} {} \times
\delta(x-x_{1})\
\psi^{\uparrow \ *}_{(n)}(x^\prime_i,
  {\vec{k}}^\prime_{\perp i},\lambda_i) \
\psi^{\uparrow}_{(n)}(x_i, {\vec{k}}_{\perp i},\lambda_i) \ ,
\nonumber \\
\lefteqn{
{1 \over \sqrt{1-\zeta}} \;
{\Delta^1-{i} \Delta^2\over 2M}\,
E_{(n\to n)}(x,\zeta,t)
\rule{0pt}{4.5ex} } \hspace{5em}
\label{t1f2}\\
 &=&
{\sqrt{1-\zeta}^{\, 2-n}}\, \sum_{n, \lambda_i}
\int \prod_{i=1}^{n}
{{\rm d}x_{i}\, {\rm d}^2{\vec{k}}_{\perp i} \over 16\pi^3 }\ \,
16\pi^3 \delta\left(1-\sum_{j=1}^n x_j\right) \, \delta^{(2)}
\left(\sum_{j=1}^n {\vec{k}}_{\perp j}\right)
\nonumber\\
&& \qquad \qquad \rule{0pt}{3ex} {} \times
\delta(x-x_{1})\
\psi^{\uparrow \ *}_{(n)}(x^\prime_i,
     {\vec{k}}^\prime_{\perp i},\lambda_i) \
\psi^{\downarrow}_{(n)}(x_i, {\vec{k}}_{\perp i},\lambda_i) \ ,
\nonumber
\end{eqnarray}
where the arguments of the final-state wavefunction are given by
\begin{equation}
\begin{array}[t]{lll}
x^\prime_{1} = {\displaystyle \frac{x_{1}-\zeta}{1-\zeta}}\, ,\
&{\vec{k}}^\prime_{\perp 1} ={\vec{k}}^{~}_{\perp 1}
- {\displaystyle \frac{1-x_{1}}{1-\zeta}}\, {\vec{\Delta}}_\perp
&\mbox{for the struck quark,}
\\[2ex]
x^\prime_i = {\displaystyle \frac{x_i}{1-\zeta}}\, ,\
&{\vec{k}}^\prime_{\perp i} ={\vec{k}}^{~}_{\perp i}
+ {\displaystyle \frac{x_i}{1-\zeta}}\, {\vec{\Delta}}_\perp
&\mbox{for the spectators $i=2, \cdots, n$.}
\end{array}
\label{t2}
\end{equation}
One easily checks that $\sum_{i=1}^n x^\prime_i = 1$ and $\sum_{i=1}^n
{\vec{k}}^\prime_{\perp i} = {\vec{0}}_\perp$. In Eqs.~(\ref{t1}) and
(\ref{t1f2}) one has to sum over all possible combinations of
helicities $\lambda^{~}_i$ and over all parton numbers $n$ in the Fock
states. We also imply a sum over all possible ways of numbering the
partons in the $n$-particle Fock state so that the struck quark has
the index $i=1$.

Analogous formulae hold in the domain $\zeta-1 < x <0$, where the
struck parton in the target is an antiquark instead of a quark. Some
care has to be taken regarding overall signs arising because fermion
fields anticommute. For details we refer to
\cite{Diehl:1999kh,Diehl:2000xz}.

For the $n+1 \to n-1$ off-diagonal term ($\Delta n = -2$), let us
consider the case where quark $1$ and antiquark $n+1$ of the initial
wavefunction annihilate into the current leaving $n-1$ spectators.
Then $x_{n+1} = \zeta - x_{1}$ and ${\vec{k}}_{\perp n+1} =
{\vec{\Delta}}_\perp-{\vec{k}}_{\perp 1}$. The remaining $n-1$ partons
have total plus-momentum $(1-\zeta)P^+$ and transverse momentum
$-{\vec{\Delta}}_{\perp}$.  The current matrix element now is
\begin{eqnarray}
\lefteqn{
\int\frac{d y^-}{8\pi}\;e^{ix P^+y^-/2}\;
\left< 0 \left| \bar\psi(0)\, {\gamma^+}\,\psi(y)\, \right|
2;\ x^{~}_{1} P^+, x^{~}_{n+1} P^+,\;
    {\vec p^{~}_{\perp 1}}, {\vec p^{~}_{\perp n+1}},\;
    \lambda^{~}_{1}, \lambda^{~}_{n+1}
\right> \Big|_{y^+=0, y_\perp=0}
} \hspace{21em}
\\
&=& \sqrt{x_{1} x_{n+1} \phantom{1}\!\!}\, \delta(x-x_{1})\
    \delta_{\lambda_{1}\, -\lambda_{n+1}}\ ,
\nonumber
\end{eqnarray}
and we thus obtain the formulae for the off-diagonal
contributions to $H$ and $E$
in the domain $0\le x \le \zeta$:
\begin{eqnarray}
\lefteqn{
{\sqrt{1-\zeta} \over 1-\zetahalf}\, H_{(n+1\to n-1)}(x,\zeta,t)\,
-\, {\zeta^2 \over 4 (1-\zetahalf) {\sqrt{1-\zeta}}} \,
    E_{(n+1\to n-1)}(x,\zeta,t)
} \hspace{5em}
\label{t3} \\
 &=&
{\sqrt{1-\zeta}^{\, 3-n}}\, \sum_{n, \lambda_i}
\int \prod_{i=1}^{n+1}
{{\rm d}x_{i}\, {\rm d}^2{\vec{k}}_{\perp i} \over 16\pi^3 }\
16\pi^3 \delta\left(1-\sum_{j=1}^{n+1} x_j\right) \, \delta^{(2)}
\left(\sum_{j=1}^{n+1} {\vec{k}}_{\perp j}\right)
\nonumber\\
&& \qquad \qquad \rule{0pt}{3ex} {} \times
16\pi^3 \delta(x_{n+1}+x_{1}-\zeta)\,
\delta^{(2)}\left( {\vec{k}}_{\perp n+1} +
{\vec{k}}_{\perp 1} - {\vec{\Delta}}_\perp \right)
\nonumber \\
&& \qquad \qquad \rule{0pt}{3ex} {} \times \delta(x-x_{1})\
\psi^{\uparrow \ *}_{(n-1)}(x^\prime_i,
  {\vec{k}}^\prime_{\perp i},\lambda^{~}_i) \
\psi^{\uparrow}_{(n+1)}(x_i, {\vec{k}}_{\perp i},\lambda_i) \
\delta_{\lambda_{1}\, -\lambda_{n+1}} \ ,
\nonumber \\
\lefteqn{
{1 \over \sqrt{1-\zeta}} \;
{\Delta^1-{i} \Delta^2\over 2M}\,
E_{(n+1\to n-1)}(x,\zeta,t)
\rule{0pt}{4.5ex} } \hspace{5em}
\label{t3f2} \\
 &=&
{\sqrt{1-\zeta}^{\, 3-n}}\, \sum_{n, \lambda_i}
\int \prod_{i=1}^{n+1}
{{\rm d}x_{i}\, {\rm d}^2{\vec{k}}_{\perp i} \over 16\pi^3 }\
16\pi^3 \delta\left(1-\sum_{j=1}^{n+1} x_j\right) \, \delta^{(2)}
\left(\sum_{j=1}^{n+1} {\vec{k}}_{\perp j}\right)
\nonumber\\
&& \qquad \qquad \rule{0pt}{3ex} {} \times
16\pi^3 \delta(x_{n+1}+x_{1}-\zeta)\,
\delta^{(2)}\left( {\vec{k}}_{\perp n+1} +
{\vec{k}}_{\perp 1} - {\vec{\Delta}}_\perp \right)
\nonumber \\
&& \qquad \qquad \rule{0pt}{3ex} {} \times \delta(x-x_{1})\
\psi^{\uparrow \ *}_{(n-1)}(x^\prime_i,
  {\vec{k}}^\prime_{\perp i},\lambda^{~}_i) \
\psi^{\downarrow}_{(n+1)}(x_i, {\vec{k}}_{\perp i},\lambda_i) \
\delta_{\lambda_{1}\, -\lambda_{n+1}} \ ,
\nonumber
\end{eqnarray}
where $i=2,\cdots ,n$ label the $n-1$ spectator partons which appear
in the final-state hadron wavefunction with
\begin{equation}
x^\prime_i = {x_i\over 1-\zeta}\, ,\qquad
{\vec{k}}^\prime_{\perp i} ={\vec{k}}^{~}_{\perp i}
+ {x_i\over 1-\zeta}\, {\vec{\Delta}}_\perp \ .
\label{t3a}
\end{equation}
We can again check that the arguments of the final-state wavefunction
satisfy $\sum_{i=2}^{n} x^\prime_i = 1$, $\sum_{i=2}^{n}
{\vec{k}}^\prime_{\perp i} = {\vec{0}}_\perp$. We imply in (\ref{t3})
and (\ref{t3f2}) a sum over all possible ways of numbering the partons
in the initial wavefunction such that the quark with index $1$ and the
antiquark with index $n+1$ annihilate into the current.

The powers of $\sqrt{1-\zeta}$ in (\ref{t1}), (\ref{t1f2}) and
(\ref{t3}), (\ref{t3f2}) have their origin in the integration measures
in the Fock state decomposition (\ref{a318}) for the outgoing
proton. The fractions $x_i'$ appearing there refer to the light-cone
momentum $P'^+ = (1-\zeta)\, P^+$, whereas the fractions $x_i$ in the
incoming proton wavefunction refer to $P^+$. Transforming all
fractions so that they refer to $P^+$ as in our final formulae thus
gives factors of $\sqrt{1-\zeta}$. Different powers appear in the
$n\to n$ and $n+1 \to n-1$ overlaps because of the different parton
numbers in the final state wavefunctions.

\section{The Virtual Compton Amplitude in QED}

\begin{figure}[htbp]
\begin{center}
\leavevmode
{\hbox{\hspace{-.7in}
\epsfbox{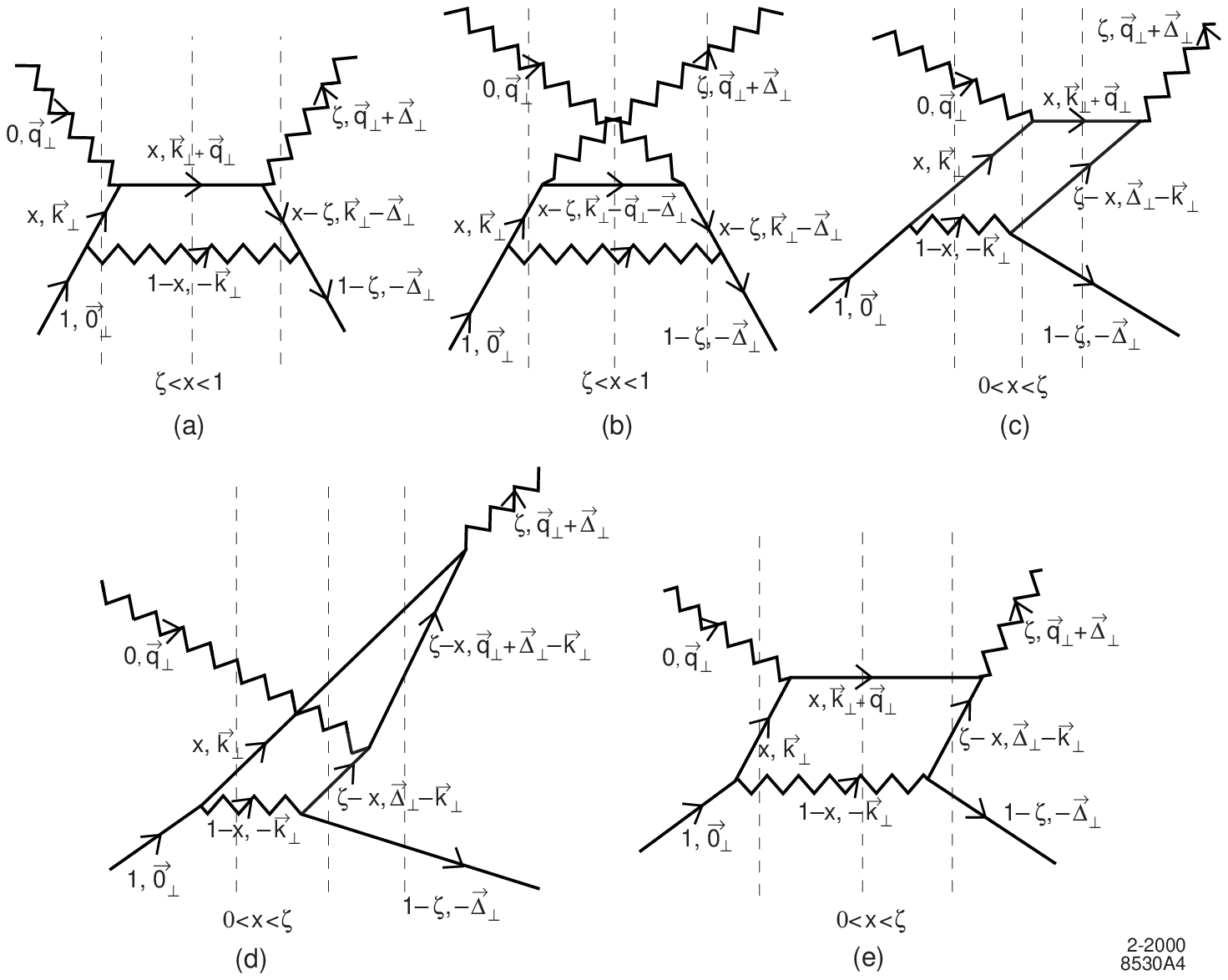}
\hss}}
\end{center}
\caption[*]{Light-cone time-ordered contributions to deeply virtual
Compton scattering on an electron in QED at one-loop order.  Note that
the contribution of figure (e) is suppressed at large $q^2$ since the
hard propagator with transverse momentum ${\vec{k}}_\perp +
{\vec{q}}_\perp$ extends over two light-cone time orderings.  The
contributions of figures (c) and (d) correspond to the overlap of
one-particle and three-particle Fock states.  The three-particle Fock
states occurs at the intermediate light-cone time indicated by the
middle vertical dashed line.}
\label{fig:5}
\end{figure}

The light-cone Fock state wavefunctions corresponding to the quantum
fluctuations of a physical electron can be systematically evaluated in
QED perturbation theory. The covariant Feynman amplitudes which
contribute to the virtual Compton amplitude at one-loop order were
illustrated in Fig.~\ref{fig:2}. The corresponding light-cone
time-ordered contributions for frames in which $q^+ = 0$ are shown in
Fig.~\ref{fig:5}.

\vspace{.5cm}

The physical electron is the eigenstate of the QED Hamiltonian.  As
discussed in Section~\ref{sec:overlap}, the expansion of the QED
eigenfunction of the electron on the complete set $\{\ket{n} \}$ of
$H_{LC}(e=0)$ eigenstates produces the Fock state expansion. Each
Fock-state wavefunction of the physical electron with total spin
projection $J^z = \pm {1\over 2}$ is represented by the function
$\psi^{J^z}_n(x_i,{\vec k}_{\perp i},\lambda_i)$, where
\begin{equation}
k_i=(k^+_i,{\vec k}_{ \perp i},k^-_i)=
\left(x_i P^+, {\vec k}^{~}_{\perp i},
\frac{{\vec k}_{\perp i}^2+m_i^2}{x_i P^+}\right)
\end{equation}
specifies the momentum of each constituent and $\lambda_i$ specifies
its light-cone helicity in the $z$ direction.

The quantum fluctuations of the electron at one-loop generate two
types of light-cone wavefunctions, $\ket{e^- \gamma}$ and $\ket{e^-
e^- e^+}$, in addition to renormalizing the one-electron state.  The
two-particle Fock state for an electron with $J^z = + {1\over 2}$ has
four possible spin combinations for the electron and photon in flight:
\begin{eqnarray}
\lefteqn{
\left|\Psi^{\uparrow}_{\rm two \ particle}(P^+, \vec P_\perp = \vec
0_\perp)\right> =
\int\frac{{\mathrm d} x \, {\mathrm d}^2
           {\vec k}_{\perp} }{\sqrt{x(1-x)}\, 16 \pi^3}
}
\label{vsn1}\\
&&
\left[ \ \ \,
\psi^{\uparrow}_{+\frac{1}{2}\, +1}(x,{\vec k}_{\perp})\,
\left| +\frac{1}{2}\, +1\, ;\,\, xP^+\, ,\,\, {\vec k}_{\perp}\right>
+\psi^{\uparrow}_{+\frac{1}{2}\, -1}(x,{\vec k}_{\perp})\,
\left| +\frac{1}{2}\, -1\, ;\,\, xP^+\, ,\,\, {\vec k}_{\perp}\right>
\right.
\nonumber\\
&&\left. {}
+\psi^{\uparrow}_{-\frac{1}{2}\, +1} (x,{\vec k}_{\perp})\,
\left| -\frac{1}{2}\, +1\, ;\,\, xP^+\, ,\,\, {\vec k}_{\perp}\right>
+\psi^{\uparrow}_{-\frac{1}{2}\, -1} (x,{\vec k}_{\perp})\,
\left| -\frac{1}{2}\, -1\, ;\,\, xP^+\, ,\,\, {\vec k}_{\perp}\right>\
\right] \ ,
\nonumber
\end{eqnarray}
where the two-particle states $|s_{\rm f}^z, s_{\rm b}^z; \ x, {\vec
k}_{\perp} \rangle$ are normalized as in (\ref{normalize}). $s_{\rm
f}^z$ and $s_{\rm b}^z$ denote the $z$-component of the spins of the
constituent fermion and boson, respectively, and the variables $x$ and
${\vec k}_{\perp}$ refer to the momentum of the fermion. The
wavefunctions can be evaluated explicitly in QED perturbation theory
using the rules given in Refs.~\cite {Lepage:1980fj,Brodsky:1980zm}:
\begin{equation}
\left
\{ \begin{array}{l}
\psi^{\uparrow}_{+\frac{1}{2}\, +1} (x,{\vec k}_{\perp})=-{\sqrt{2}}
\ \frac{-k^1+{i} k^2}{x(1-x)}\,
\varphi \ ,\\
\psi^{\uparrow}_{+\frac{1}{2}\, -1} (x,{\vec k}_{\perp})=-{\sqrt{2}}
\ \frac{k^1+{i} k^2}{1-x }\,
\varphi \ ,\\
\psi^{\uparrow}_{-\frac{1}{2}\, +1} (x,{\vec k}_{\perp})=-{\sqrt{2}}
\ (M-{m\over x})\,
\varphi \ ,\\
\psi^{\uparrow}_{-\frac{1}{2}\, -1} (x,{\vec k}_{\perp})=0\ ,
\end{array}
\right.
\label{vsn2}
\end{equation}
where
\begin{equation}
\varphi (x,{\vec k}_{\perp}) = \frac{e}{\sqrt{1-x}}\
\frac{1}{M^2-{{\vec k}_{\perp}^2+m^2 \over x}
-{{\vec k}_{\perp}^2+\lambda^2 \over 1-x}}\ .
\label{wfdenom}
\end{equation}
Similarly, the wavefunctions for an electron with negative helicity
are given by
\begin{equation}
\left
\{ \begin{array}{l}
\psi^{\downarrow}_{+\frac{1}{2}\, +1} (x,{\vec k}_{\perp})=0\ ,\\
\psi^{\downarrow}_{+\frac{1}{2}\, -1} (x,{\vec k}_{\perp})=-{\sqrt{2}}
\ (M-{m\over x})\, \varphi \ ,\\
\psi^{\downarrow}_{-\frac{1}{2}\, +1} (x,{\vec k}_{\perp})=-{\sqrt{2}}
\ \frac{-k^1+{i} k^2}{1-x }\,
\varphi \ ,\\ \psi^{\downarrow}_{-\frac{1}{2}\, -1} (x,{\vec
k}_{\perp})=-{\sqrt{2}}
\ \frac{k^1+{i} k^2}{x(1-x)}\, \varphi \ .
\end{array}
\right.
\label{vsn2a}
\end{equation}

In (\ref{vsn2}) and (\ref{vsn2a}) we have generalized the framework of
QED by assigning a mass $M$ to the external electrons in the Compton
scattering process, but a different mass $m$ to the internal electron
lines and a mass $\lambda$ to the internal photon line
\cite{Brodsky:1980zm}. The idea behind this is to model the structure
of a composite fermion state with mass $M$ by a fermion and a vector
constituent with respective masses $m$ and $\lambda$.

In the domain $\zeta < x < 1$, for a general value of $\zeta$ between
0 and 1, we have
\begin{eqnarray}
\lefteqn{
{\sqrt{1-\zeta} \over 1-\zetahalf}\ H_{(2\to 2)}(x,\zeta,t)\,
-\, {\zeta^2 \over 4 (1-\zetahalf){\sqrt{1-\zeta}}} \
E_{(2\to 2)}(x,\zeta,t) }
\label{gf3} \\
&=&
\int\frac{{\mathrm d}^2 {\vec k}_{\perp} }{16 \pi^3}
\Big[ \psi^{\uparrow\ *}_{+\frac{1}{2}\, +1}(x',{\vec k'}_{\perp})
\psi^{\uparrow}_{+\frac{1}{2}\, +1}(x,{\vec k}_{\perp})
+\psi^{\uparrow\ *}_{+\frac{1}{2}\, -1}(x',{\vec k'}_{\perp})
\psi^{\uparrow}_{+\frac{1}{2}\, -1}(x,{\vec k}_{\perp})
\nonumber\\
&&\qquad\qquad\qquad\qquad
{}+\psi^{\uparrow\ *}_{-\frac{1}{2}\, +1}(x',{\vec k'}_{\perp})
\psi^{\uparrow}_{-\frac{1}{2}\, +1}(x,{\vec k}_{\perp})
\Big]\ ,
\nonumber \\
\lefteqn{
{1 \over \sqrt{1-\zeta}} \;
{(\Delta^1-{i} \Delta^2)\over 2M}\ E_{(2\to 2)}(x,\zeta,t)
}
\label{gf4} \\
&=&
\int\frac{{\mathrm d}^2 {\vec k}_{\perp} }{16 \pi^3}
\Big[\psi^{\uparrow\ *}_{+\frac{1}{2}\, -1}(x',{\vec k'}_{\perp})
\psi^{\downarrow}_{+\frac{1}{2}\, -1}(x,{\vec k}_{\perp})
+\psi^{\uparrow\ *}_{-\frac{1}{2}\, +1}(x',{\vec k'}_{\perp})
\psi^{\downarrow}_{-\frac{1}{2}\, +1}(x,{\vec k}_{\perp})
\Big] \ ,
\nonumber
\end{eqnarray}
where
\begin{equation}
x'={x-\zeta\over 1-\zeta},\ \ \
{\vec k'}_{\perp}={\vec k}^{~}_{\perp}-{1-x\over
1-\zeta}\ {\vec{\Delta}}_{\perp}\ .
\label{xprime}
\end{equation}
These contributions correspond to the overlap of the two-particle Fock
components of the electron as illustrated in Figs.~\ref{fig:5}(a) and
\ref{fig:5}(b). The generalized form factors $H_{(2\to 2)}(x,\zeta,t)$
and $E_{(2\to 2)}(x,\zeta,t)$ are zero in the domain $\zeta-1 < x <
0$, which corresponds to emission and reabsorption of an $e^+$ from a
physical electron. Contributions to $H_{(n\to n)}(x,\zeta,t)$ and
$E_{(n\to n)}(x,\zeta,t)$ in that domain only appear beyond one-loop
level.

At this point, a comment is in order on the large ${\vec k}_{\perp}$
behavior of the overlap integrals (\ref{gf3}) and (\ref{gf4}). They
are logarithmically divergent, which reflects the fact that the matrix
elements defining parton distributions have to be regulated and
renormalized in the ultraviolet. The dependence of parton
distributions on the renormalization scale can be calculated
perturbatively and is expressed in the well-known evolution equations.
A physically intuitive way to implement this in our context is to
introduce an upper cutoff in the invariant mass of the Fock states
\cite{Lepage:1980fj}, which roughly speaking corresponds to an upper
cutoff in the transverse parton momenta. How this is to be done in
detail, and how it leads to the evolution equations for the
generalized Compton form factors
\cite{Muller:1994fv,Ji:1997ek,Radyushkin:1997ki} is beyond the scope
of the present paper.

We will also require three-constituent wavefunctions corresponding to
two electrons and one positron in flight:
\begin{equation}
\psi^{J^z}_{_{s^z_1\,\, s^z_2\,\, s^z_3}}
(x_1,x_2,x_3,{\vec k}_{1\perp},{\vec k}_{2\perp},{\vec k}_{3\perp})=
M^{J^z}_{s^z_1\,\, s^z_2\,\, s^z_3}\
{\widetilde{\varphi}}
(x_1,x_2,x_3,{\vec k}_{1\perp},{\vec k}_{2\perp},{\vec k}_{3\perp}),
\label{psi3}
\end{equation}
where according to our numbering convention spelled out after
(\ref{t3a}) the index $1$ refers to the electron with mass $m$, the
index $2$ to the electron with mass $M$, and the index $3$ to the
positron. The denominator part of the wavefunction
${\widetilde{\varphi}}$ is given by
\begin{eqnarray}
\lefteqn{
{\widetilde{\varphi}}
(x_1,x_2,x_3,{\vec k}_{1\perp},{\vec k}_{2\perp},{\vec k}_{3\perp})
= {e^2  \over 1-x_1}
}
\label{wphi}\\
&& {}\times
{1\over \left(M^2-{m^2+{\vec{k}}_{1\perp}^2\over x_1}
-{\lambda^2+{\vec{k}}_{1\perp}^2\over 1-x_1}\right)
\left(M^2-{m^2+{\vec{k}}_{1\perp}^2\over x_1}
-{M^2+{\vec{k}}_{2\perp}^2\over x_2}
-{m^2+{\vec{k}}_{3\perp}^2\over x_3}
\right)}\ .
\nonumber
\end{eqnarray}
The numerator factors $M^{J^z}_{s^z_1\,\, s^z_2\,\, s^z_3}$ in
(\ref{psi3}) in QED are given in Table 1.

\begin{figure}[htb]
\begin{center}
\leavevmode
{
 \epsfxsize=4.75in\epsfbox{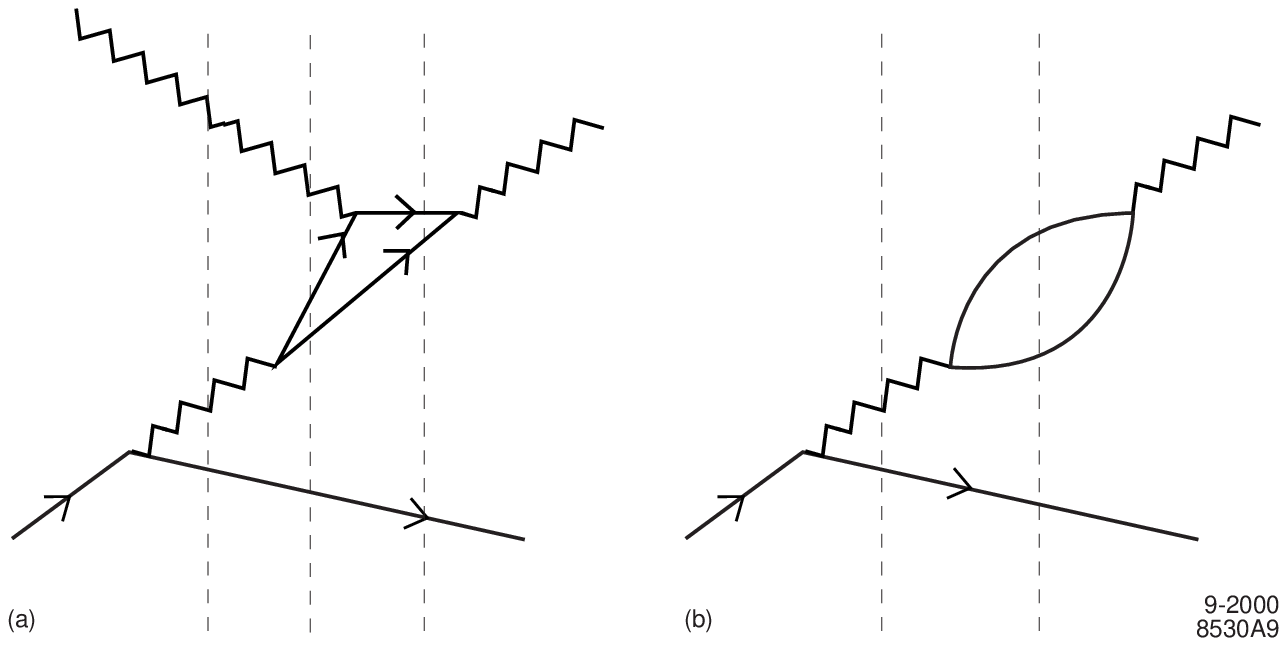}
}\end{center}
\caption[*]{Light-cone time-ordered diagrams where the electron
coupling to the external current originates from the splitting of an
intermediate photon. Diagram (a) for deeply virtual Compton scattering
vanishes because of Furry's theorem.
The corresponding diagram (b) for the
electromagnetic vertex gives the vacuum polarization correction for the
external photon.}
\label{fig:5a}
\end{figure}

In the wave functions (\ref{psi3}) the positron and the electron with
index $2$ originate in the splitting of the intermediate photon,
corresponding to the diagrams of Fig~\ref{fig:5}(c) and (d). We note
that there is also a contribution to the three-particle wavefunctions
where the photon splits into the positron and the electron with index
$1$. These terms do not contribute to deeply virtual Compton
scattering because of charge conjugation invariance; see
Fig.~\ref{fig:5a}(a). They do contribute in the calculation of the
electromagnetic vertex as shown in Fig.~\ref{fig:5a}(b), where they
provide the vacuum polarization correction for the external
photon. This correction is usually excluded in the definition of the
electromagnetic form factors $F_1$ and $F_2$, and thus of the first
moments of $H$ and $E$. We will therefore discard the corresponding
contributions to $H(x,\zeta,t)$ and $E(x,\zeta,t)$ in the following.

\vspace{.5cm}

\begin{table}[htb]
\begin{center}
Table 1. Numerator factors of the three-constituent wavefunctions in
QED. We only list those helicity combinations for which $s_1^z = -
s_3^z$. \\
\vspace{.6truecm}
\begin{footnotesize}
\begin{tabular}{|c|c|c|}
\hline\ru1
$\ \ \ \ J^z\ \ \ \ $ & $ \ \ s^z_1\ \ \ \ s^z_2\ \ \ \ s^z_3\ \ $
&$\ \ \ {1\over 2}\ M^{J^z}_{s^z_1\,\, s^z_2\,\, s^z_3}\ \ \ $\\
\hline\hline\ru1
$+{1\over 2}$
& $\ \ +{1\over 2}\ \ +{1\over 2}\ \ -{1\over 2}\ \ $
& $ -\frac{k_1^1-{i} k_1^2}{x_1(1-x_1)}\,
({k_3^1+{i}k_3^2\over x_3}+{k_1^1+{i}k_1^2\over 1-x_1})
\, - \, \frac{k_1^1+{i} k_1^2}{1-x_1 }\,
({k_2^1-{i}k_2^2\over x_2}+{k_1^1-{i}k_1^2\over 1-x_1})$ \\
\hline\ru1
$+{1\over 2}$
& $\ \ -{1\over 2}\ \ +{1\over 2}\ \ +{1\over 2}\ \ $
&$ -(M-{m\over x_1})\, ({M\over x_2}+{m\over x_3})$ \\
\hline\ru1
$+{1\over 2}$
& $\ \ +{1\over 2}\ \ -{1\over 2}\ \ -{1\over 2}\ \ $
&$-\frac{k_1^1+{i} k_1^2}{1-x_1 }\,
({M\over x_2}+{m\over x_3})$ \\
\hline\ru1
$+{1\over 2}$
& $\ \ -{1\over 2}\ \ -{1\over 2}\ \ +{1\over 2}\ \ $
& $ (M-{m\over x_1})\,
({ k_2^1+{i}k_2^2\over x_2}+{ k_1^1+{i}k_1^2\over 1-x_1})$ \\
\hline\hline\ru1
$-{1\over 2}$
& $\ \ +{1\over 2}\ \ +{1\over 2}\ \ -{1\over 2}\ \ $
& $-(M-{m\over x_1})\,
({k_2^1-{i}k_2^2\over x_2}+{k_1^1-{i}k_1^2\over 1-x_1})$ \\
\hline\ru1
$-{1\over 2}$
& $\ \ -{1\over 2}\ \ +{1\over 2}\ \ +{1\over 2}\ \ $
&$ \frac{k_1^1-{i} k_1^2}{1-x_1 }\,
({M\over x_2}+{m\over x_3})$ \\
\hline\ru1
$-{1\over 2}$
& $\ \ +{1\over 2}\ \ -{1\over 2}\ \ -{1\over 2}\ \ $
&$-(M-{m\over x_1})\, ({M\over x_2}+{m\over x_3})$ \\
\hline\ru1
$-{1\over 2}$
& $\ \ -{1\over 2}\ \ -{1\over 2}\ \ +{1\over 2}\ \ $
& $-\frac{k_1^1+{i} k_1^2}{x_1(1-x_1)}\,
({k_3^1-{i}k_3^2\over x_3}+{k_1^1-{i}k_1^2\over 1-x_1})
\, - \, \frac{k_1^1-{i} k_1^2}{1-x_1 }\,
({k_2^1+{i}k_2^2\over x_2}+{k_1^1+{i}k_1^2\over 1-x_1})$ \\
\hline
\end{tabular}
\end{footnotesize}
\end{center}
\end{table}

The electron in QED also has a one-particle component:
\begin{equation}
\left|\Psi_{\rm one \ particle}^{\uparrow , \downarrow}
(P^+, \vec P_\perp = \vec 0_\perp)\right> =
\int {{\mathrm d}x\, {\mathrm d}^2 {\vec{k}}_{\perp} \over
\sqrt{x}\, 16\pi^3}\ 16\pi^3 \delta (1-x)\,
{\delta}^2({\vec{k}}_{\perp})\
\psi_{(1)}\ \left| \pm {1\over 2} \, ;
xP^+, {\vec{k}}_{\perp} \right>
\label{bare1}
\end{equation}
where the one-constituent wavefunction is given by
\begin{equation}
\psi_{(1)} = \sqrt{Z} .
\label{oneparticle}
\end{equation}
Here $\sqrt Z$ is the wavefunction renormalization of the one-particle
state and ensures overall probability conservation. Since we are
working consistently to ${\cal O} (\alpha) $, we can set $Z = 1$ in
the $3 \to 1$ wavefunction overlap contributions.  In the domain $0 <
x < \zeta$, we then have
\begin{eqnarray}
\lefteqn{
{\sqrt{1-\zeta} \over 1-\zetahalf}\ H_{(3\to 1)}(x,\zeta,t)\,
-\, {\zeta^2 \over 4 (1-\zetahalf){\sqrt{1-\zeta}}}\
E_{(3\to 1)}(x,\zeta,t)
}
\label{gf331}\\
&=&
\sqrt{1-\zeta}\,
\int {{\rm d}^2{\vec{k}}_{\perp}\over 16\pi^3}\
\left[ \psi^{\uparrow}_{_{+{1\over 2}\,\,
+{1\over 2}\,\, -{1\over 2}}}
(x,1-\zeta,\zeta -x,\ {\vec k}_{\perp}, -{\vec \Delta}_{\perp},
{\vec \Delta}_{\perp}-{\vec k}_{\perp})
\right.
\nonumber
\\
&& \hspace{6.4em} \left. {} +
\psi^{\uparrow}_{_{-{1\over 2}\,\, +{1\over 2}\,\, +{1\over 2}}}
(x,1-\zeta,\zeta -x,{\vec k}_{\perp},-{\vec \Delta}_{\perp},
{\vec \Delta}_{\perp}-{\vec k}_{\perp})
\right]\ ,
\nonumber \\
\lefteqn{
{1 \over \sqrt{1-\zeta}} \;
{(\Delta^1-{i} \Delta^2)\over 2M}\
E_{(3\to 1)}(x,\zeta,t)
}
\label{gf431}\\
&=&
\sqrt{1-\zeta}\,
\int {{\rm d}^2{\vec{k}}_{\perp}\over 16\pi^3}\
\left[ \psi^{\downarrow}_{_{+{1\over 2}\,\,
+{1\over 2}\,\, -{1\over 2}}}
(x,1-\zeta,\zeta -x,\ {\vec k}_{\perp}, -{\vec \Delta}_{\perp},
{\vec \Delta}_{\perp}-{\vec k}_{\perp})
\right.
\nonumber
\\
&& \hspace{6.4em} \left. {} +
\psi^{\downarrow}_{_{-{1\over 2}\,\, +{1\over 2}\,\, +{1\over 2}}}
(x,1-\zeta,\zeta -x,{\vec k}_{\perp},-{\vec \Delta}_{\perp},
{\vec \Delta}_{\perp}-{\vec k}_{\perp})
\right]\ .
\nonumber
\end{eqnarray}
These contributions correspond to the overlap of the three-particle
and one-particle Fock components as illustrated in
Figs.~\ref{fig:5}(c) and \ref{fig:5}(d).

\begin{figure}[htb]
{\hspace{-1in}
\begin{center}
\leavevmode
 \epsfbox{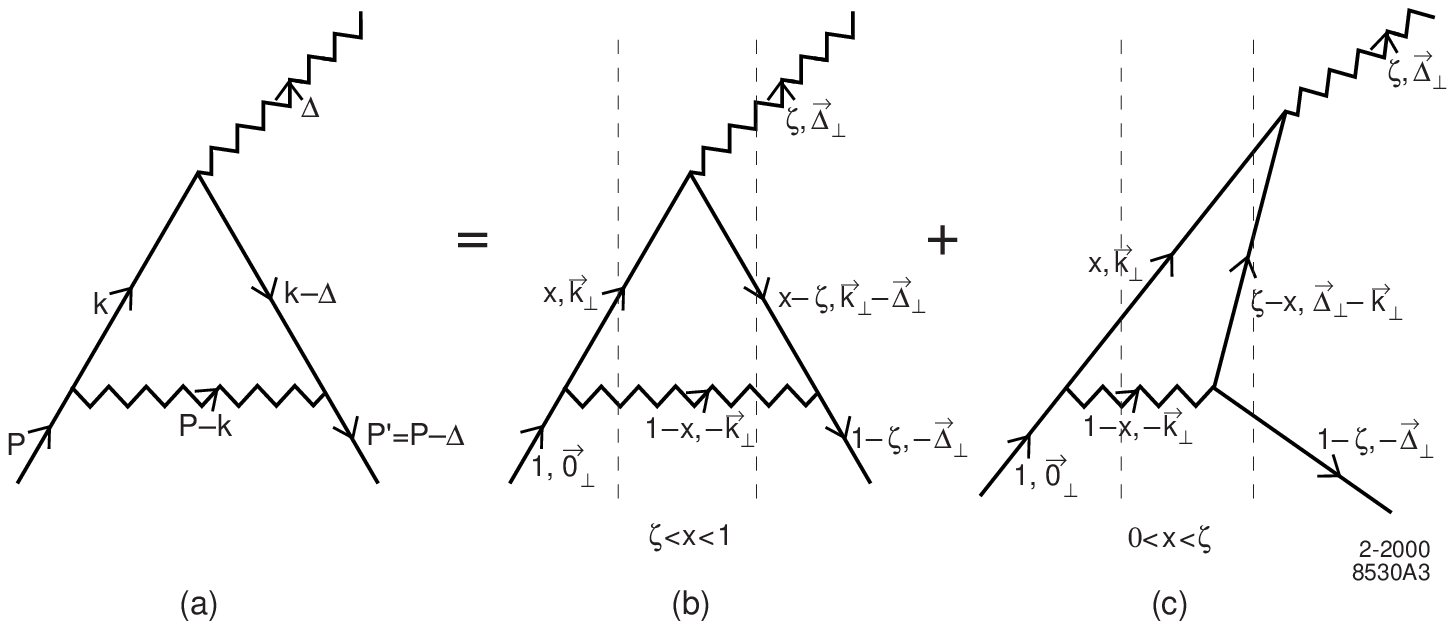}
\end{center}}
\caption[*]{Light-cone time-ordered contributions to the form factors
of an electron in QED at one-loop order. The spacelike form factors
are independent of the choice of $\zeta$ at fixed $t = \Delta^2$. In
particular, if $\zeta = 0$ the contribution of amplitude (c) to the
matrix elements of $J^+(0)$ vanishes.}
\label{fig:6}
\end{figure}

The first moments (\ref{d1}) and (\ref{d3}) of the generalized Compton
form factors at one loop give the one-loop space-like form factors
$F_1$ and $F_2$. The corresponding light-cone time-ordered diagrams
are shown in Fig.~\ref{fig:6}. As in our simple example discussed at
the end of Section~\ref{sec:overlap}, the sum of the two diagrams must
give the same result as the corresponding Feynman diagram in covariant
perturbation theory. The division of the integral over $x$ in the form
factors (\ref{d1}) and (\ref{d3}) into contributions from $n\to n$ and
$n+1\to n-1$ transitions is again a consequence of an arbitrary choice
of frame and of light-cone direction, and the sum of the contributions
is thus independent of the light-cone variable $\zeta$.

We also note that the integrands in (\ref{d1}) and (\ref{d3}) have to
be continuous at the point $x=\zeta$ which separates $n\to n$ and
$n+1\to n-1$ transitions. In other words, the generalized form factors
$H(x,\zeta,t)$ and $E(x,\zeta,t)$ must be continuous functions of~$x$
at $x=\zeta$. This is required for the loop integral (\ref{com1a}) of
the Compton amplitude to exist, given the form (\ref{com2p}) of the
hard scattering subprocess. The continuity of $H$ and $E$ in one-loop
QED can readily be checked from our results~(\ref{gf3}), (\ref{gf4}),
(\ref{gf331}), (\ref{gf431}). The underlying relations between the
one, two, and three particle wavefunctions of the electron reflect
again the Lorentz frame invariance of the light-cone Fock state
representation.

\vspace{.5cm}

\section {Conclusions}
\label{sec:con}

A central goal of quantum chromodynamics is to determine the structure
of hadrons in terms of their quark and gluon degrees of freedom.  As
we have shown in this paper, the deeply virtual Compton exclusive
process $\gamma^* p \to \gamma p$ provides a direct window into hadron
substructure which goes well beyond inclusive measures.

The deeply virtual Compton amplitude has a simple representation in
terms of the light-cone Fock wavefunctions of the target, factorizing
as the convolution of a hard perturbative amplitude, corresponding to
Compton scattering on a quark current, with the initial and final
state light-cone wavefunctions of the target hadron. The light-cone
Fock representation provides an explicit and physical representation
of the leading-twist operator product decomposition for the deeply
virtual Compton amplitude.  As in the case of time-like semi-leptonic
decays of hadrons \cite{Brodsky:1999hn}, there are two distinct
contributions: a parton-number conserving diagonal overlap integral of
light-cone wavefunctions, plus an additional $\Delta n = -2$
contribution where the quark-antiquark pair of the initial state is
annihilated.

The light-cone Fock representation also provides a direct derivation
of the identity between the form factor densities $H(x,\zeta,t)$ and $
E(x,\zeta,t)$ which appear in deeply virtual Compton scattering and
the corresponding integrands of the Dirac and Pauli form factors
$F_{1}(t)$ and $F_{2}(t)$, and the gravitational form factors
$A_{q}(t)$ and $B_{q}(t)$ for each quark and anti-quark
constituent. Thus deeply virtual Compton scattering effectively
provides access to the form factors of a proton scattering in a
gravitational field.

A remarkable feature of these sum rules is the fact that the
integrals over $x$ of $(1-\zeta/2)^{-1}\, H(x,\zeta,t)$ and
$(1-\zeta/2)^{-1}\, E(x,\zeta,t)$ are independent of the value of
$\zeta$. This invariance is due to the Lorentz frame-independence of
the light-cone Fock representation of space-like local operator matrix
elements. This frame independence in turn reflects the underlying
connections between Fock states of different parton number implied by
the QCD equations of motion.

We have illustrated our general formalism by computing deeply virtual
Compton scattering on the quantum fluctuations of a fermion in QED at one
loop.  These forms can be simply generalized using Pauli-Villars spectral
integrals to provide a self-consistent model of hadron structure.  Such a
model builds in all of the constraints of Lorentz invariance, including
$J^z$ conservation and the required connections of the $n$, $n-1$, and
$n+1$-particle Fock states.  Such a model thus provides the simplest
possible template for parametrizing and interrelating hadronic structure
as measured by form factors, deep inelastic scattering and deeply virtual
Compton scattering.

\section*{Appendix: Formulae in the Symmetric Frame}

It is often convenient to choose a ``symmetric'' light-cone frame for
the momenta of the initial and final target proton which has $\Delta
\to -\Delta$ symmetry. In this frame, one parametrizes the initial
and final target momenta as \cite{Ji:1997ek}:
\begin{equation}
P=\left( (1 +\xi) {\bar P}^+\ ,\ {\vec \Delta_\perp/2}\ ,\ {M^2+{\vec
\Delta_\perp}^2/4 \over (1+\xi){\bar P}^+}\right)\ ,
\label{app1}
\end{equation}
\begin{equation}
P'= \left((1- \xi){\bar P}^+\ ,\ -{ \vec \Delta_\perp/2}\ ,\ {M^2+{\vec
\Delta_\perp}^2/4 \over (1-\xi){\bar P}^+}\right)\ .
\label{app2}
\end{equation}
The four-momentum transfer from the target then reads
\begin{equation}
\Delta\ =
\left( 2\xi {\bar P}^+\ ,\ {\vec \Delta_\perp}\ ,\
{(t+{\vec \Delta_\perp}^2)\over 2\xi {\bar P}^+}\right)\ ,
\label{app3}
\end{equation}
and one has
\begin{equation}
t= - {4\xi^ 2M^2+{\vec \Delta_\perp}^2 \over 1\ -\ \xi^2}\ .
\label{app5}
\end{equation}
Notice that our definition of the transfer $\Delta$ has the opposite
sign of Ji's.
Again we choose a light-cone frame where the incident space-like
photon carries \mbox{$q^+ = 0$}:
\begin{eqnarray}
q &=& \left( 0\ ,\ {\vec q_\perp}\ ,\
{({\vec q_\perp}+{\vec \Delta_\perp})^2\over 2\xi {\bar P}^+}
+{2\xi (M^2+{\vec \Delta_\perp}^2/4)\over
  (1-\xi^2){\bar P}^+}\right)\ ,
\label{app4}\\
q' &=&
\left( 2\xi {\bar P}^+\ ,\ {\vec q_\perp}+{\vec \Delta_\perp}\ ,\
{({\vec q_\perp}+{\vec \Delta_\perp})^2\over 2\xi {\bar P}^+}\right)\ .
\nonumber
\end{eqnarray}
In the same way as in Section~\ref{sec:kinematics} one can relate $\xi$
to the invariants of the problem. Taking the limit of large $Q^2$ at
small $t$ and comparing with (\ref{nn3}) we obtain the relation
\begin{equation}
\zeta = {2\xi \over 1+\xi}\ .
\label{new-xi}
\end{equation}
We remark that the symmetric frame just introduced and the one
described by (\ref{a1}) to (\ref{a2p}) are related by a transverse
boost. One finds that the transverse components of $\Delta$ in the two
frames are related by
\begin{equation}
\vec \Delta_\perp \Big|_{\mathrm Eq.~(\protect\ref{delta})}
= {1 \over 1+\xi }\
\vec \Delta_\perp \Big|_{\mathrm Eq.~(\protect\ref{app3})}
\ .
\label{new-delta}
\end{equation}

The deeply virtual Compton amplitude can now be written as
\begin{eqnarray}
\lefteqn{
M^{IJ}({\vec q_\perp},{\vec \Delta_\perp},\zeta)\ =\
- e^2_{q}\ {1 \over 2\bar P^+}
\int_{-1}^1d{\bar x}\
}
\label{com1a-new}
\\
&\times& \left\{ \ {\bar t}^{IJ}({\bar x},\xi)\ {\bar U}(P')
\left[
H({\bar x},\xi,t)\ {\gamma^+}
 +
E({\bar x},\xi,t)\
{i\over 2M}\, {\sigma^{+\alpha}}(-\Delta_\alpha)
\right] U(P) \right.
\nonumber\\
&& \ \
\left. {\bar s}^{IJ}({\bar x},\xi)\ {\bar U}(P')
\left[
{\widetilde H}({\bar x},\xi,t)\ {\gamma^+\gamma_5}
 +
{\widetilde E}({\bar x},\xi,t)\
{1\over 2M}\, {\gamma_5}(-\Delta^+)
\right] U(P)\
\right\}
\nonumber
\end{eqnarray}
with
\begin{eqnarray}
{\bar t}^{\ \uparrow\uparrow}({\bar x},\xi)&=&
\phantom{-} \ {\bar t}^{\ \downarrow\downarrow}({\bar x},\xi)\ =\
{1\over {\bar x}+\xi-i\epsilon}\
+\ {1\over {\bar x}-\xi +i\epsilon}\ ,
\label{com2p-new}\\
{\bar s}^{\ \uparrow\uparrow}({\bar x},\xi)&=&
-\ {\bar s}^{\ \downarrow\downarrow}({\bar x},\xi)\ =\
{1\over {\bar x}+\xi-i\epsilon}\ -\ {1\over {\bar x}-\xi +i\epsilon}\ ,
\nonumber\\
{\bar t}^{\ \uparrow\downarrow}({\bar x},\xi)
&=& {\bar t}^{\ \downarrow\uparrow}({\bar x},\xi)
\ =\ {\bar s}^{\ \uparrow\downarrow}({\bar x},\xi)
\ =\ {\bar s}^{\ \downarrow\uparrow}({\bar x},\xi)
\ \ =\ \ 0\ \phantom{\frac{1}{2}}.
\nonumber
\end{eqnarray}
for circularly polarized photons. The variable $\bar x$ is related to
$x$ in Section~\ref{sec:form-factors} by
\begin{equation}
x = \frac{\bar x + \xi}{1+\xi}
\label{new-x}
\end{equation}
and is again chosen such as to make symmetry relations under
$\Delta\to -\Delta$ most transparent.

Using the transformation rules (\ref{new-xi}), (\ref{new-delta}), and
(\ref{new-x}) it is straightforward to translate all our results into
the variables in the symmetric frame. For convenience, we give in the
following our main formulae explicitly. The spinor products in
(\ref{com1z}) and (\ref{ab1}) now read
\begin{eqnarray}
{1 \over 2\bar P^+}\
{\bar U}(P',{\lambda}'){\gamma^+}\, U(P,\lambda) &=&
{\sqrt{1-\xi^2}}\ \delta_{\lambda ,\, {\lambda}'}\ ,
\\
{1 \over 2\bar P^+}\
{\bar U}(P',{\lambda}'){i \over 2M}\,
{\sigma^{+\alpha}}(-\Delta_\alpha) U(P,\lambda)
&=&
-\, {\xi^2 \over \sqrt{1-\xi^2}}\
    \delta_{\lambda ,\, {\lambda}'}
\nonumber\\
&& +\
{1 \over \sqrt{1-\xi^2}} \;
{\, -\lambda \Delta^1\, -\, i\Delta^2\, \over 2M}\
\delta_{\lambda ,\, -{\lambda}'}\ ,
\nonumber \\
{1 \over 2M}\ {\bar U}(P',{\lambda}')U(P,\lambda) &=&
{1 \over {\sqrt{1-\xi^2}}}\ \delta_{\lambda ,\, {\lambda}'}
\nonumber \\
&& -{1\over {\sqrt{1-\xi^2}}} \;
{-\lambda \Delta^1\, -\, i\Delta^2 \over 2M}\
\delta_{\lambda ,\, -{\lambda}'}\ ,
\end{eqnarray}
and the sum rules for the form factors take the simple forms
\begin{eqnarray}
\int_{-1}^1 d{\bar x}\ H(\bar x,\xi,t) &=& F_{1}(t) \ ,
\\
\int_{-1}^1 d{\bar x}\ E(\bar x,\xi,t) &=& F_{2}(t) \ ,
\nonumber \\
\int_{-1}^1 d{\bar x}\, {\bar x} \,
   [\, H(\bar x,\xi,t) + E(\bar x,\xi,t)\, ] &=&
A_{q}(t) + B_{q}(t)  \ .
\nonumber
\end{eqnarray}

The Fock state representations of the Compton form factors $H$ and $E$
are again more symmetric with respect to the initial and final state
proton if we use the variables of the symmetric frame, although at the
price of somewhat more involved relations between the different
momentum variables. For the $n \to n$ diagonal term ($\Delta n = 0$)
we obtain in the domain $\xi \le {\bar x}\le 1$:
\begin{eqnarray}
\lefteqn{
{\sqrt{1-\xi^2}}\ H_{(n\to n)}(\bar{x},\xi,t)\,
-\, {\xi^2 \over {\sqrt{1-\xi^2}}}\,
E_{(n\to n)}(\bar{x},\xi,t)
}
\label{t1app}\\
&=&
\sqrt{1-\xi}^{\, 2-n} \sqrt{1+\xi}^{\, 2-n}\,
\sum_{n, \lambda_i}
\int
\prod_{i=1}^{n}
{{\rm d}x_{i}\, {\rm d}^2{\vec{k}}_{\perp i} \over 16\pi^3 }\
16\pi^3 \delta\left(1-\sum_{j=1}^n x_j\right)\, \delta^{(2)}
\left(\sum_{j=1}^n {\vec{k}}_{\perp j}\right)  \nonumber\\[1ex]
&&\times
\delta({\bar{x}}-x_1)\
\psi^{\uparrow \ *}_{(n)}(x^\prime_i, {\vec{k}}^\prime_{\perp i},\lambda_i)
~ \psi^{\uparrow}_{(n)}(y_i, {\vec{l}}_{\perp i},\lambda_i),
\nonumber\\[1ex]
\lefteqn{
{1\over {\sqrt{1-\xi^2}}}\,
{\Delta^1-{i} \Delta^2\over 2M}\
E_{(n\to n)}(\bar{x},\xi,t)
}
\label{t1f2app}\\
&=&
\sqrt{1-\xi}^{\, 2-n} \sqrt{1+\xi}^{\, 2-n}\,
\sum_{n, \lambda_i}
\int
\prod_{i=1}^{n}
{{\rm d}x_{i}\, {\rm d}^2{\vec{k}}_{\perp i} \over 16\pi^3 }\
16\pi^3 \delta\left(1-\sum_{j=1}^n x_j\right)\, \delta^{(2)}
\left(\sum_{j=1}^n {\vec{k}}_{\perp j}\right)  \nonumber\\[1ex]
&&\times
\delta({\bar{x}}-x_1)\
\psi^{\uparrow \ *}_{(n)}(x^\prime_i, {\vec{k}}^\prime_{\perp i},\lambda_i)
~ \psi^{\downarrow}_{(n)}(y_i, {\vec{l}}_{\perp i},\lambda_i),
\nonumber
\end{eqnarray}
where
\begin{equation}
\begin{array}[t]{lll}
x^\prime_1 = \displaystyle \frac{{x_1}-\xi}{1-\xi}\, ,\
&{\vec{k}}^\prime_{\perp 1} ={\vec{k}}_{\perp 1}
- \displaystyle \frac{1-x_1}{1-\xi}\, \frac{\vec\Delta_\perp}{2}
&\mbox{for the final struck quark,}\\[1ex]
x^\prime_i = \displaystyle \frac{x_i}{1-\xi}\, ,\
&{\vec{k}}^\prime_{\perp i} ={\vec{k}}_{\perp i}
+ \displaystyle \frac{x_i}{1-\xi}\, \frac{\vec\Delta_\perp}{2}
&\mbox{for the final $ (n-1)$ spectators,}
\end{array}
\label{t2app}
\end{equation}
and
\begin{equation}
\begin{array}[t]{lll}
y_1 = \displaystyle \frac{x_1+\xi}{1+\xi}\, ,\
&{\vec{l}}_{\perp 1} ={\vec{k}}_{\perp 1}
+ \displaystyle \frac{1-x_1}{1+\xi}\, \frac{\vec\Delta_\perp}{2}
&\mbox{for the initial struck quark,}\\[1ex]
y_i = \displaystyle \frac{x_i}{1+\xi}\, ,\
&{\vec{l}}_{\perp i} ={\vec{k}}_{\perp i}
- \displaystyle \frac{x_i}{1+\xi}\, \frac{\vec\Delta_\perp}{2}
&\mbox{for the initial $ (n-1)$ spectators.}
\end{array}
\label{t2app2}
\end{equation}
We can again check that $\sum_{i=1}^n x^\prime_i = 1$ and
$\sum_{i=1}^n {\vec{k}}^\prime_{\perp i} = {\vec{0}}_\perp$.  We also
have $\sum^{n}_{i=1} y_i =1 $ and $\sum^{n}_{i=1} {\vec{l}}_{\perp i}
={\vec{0}}_\perp$ as required. {}From (\ref{t2app}) and (\ref{t2app2})
we see that the variable $\bar{x}=x_1$ corresponds to the average
momentum fraction $(y_1 P^+ + x'_1 P'^+) /(P^+ + P'^+)$ of the struck
quark before and after the scattering.

For the $n+1 \to n-1$ off-diagonal term ($\Delta n = -2$), let us
consider the case where partons $1$ and $n+1$ of the initial
wavefunction annihilate into the current leaving $n-1$ spectators. The
final state $n-1$ parton wavefunction then has arguments $(i = 2,
\cdots ,n)$
\begin{equation}
x^\prime_i = {x_i\over 1-\xi}\ ,\qquad
{\vec{k}}^\prime_{\perp i} = {\vec{k}}_{\perp i} +
{x_i\over 1-\xi}\frac{\vec\Delta_\perp}{2} \ .
\label{appn1}
\end{equation}
We can check that $\sum_{i=2}^n x^\prime_i = 1$ and $\sum_{i=2}^n
{\vec{k}}^\prime_{\perp i} = {\vec{0}}_\perp$. The initial state $n
+1$ parton wavefunction has arguments $(i = 1, \cdots , n + 1)$,
\begin{eqnarray}
y_1&=&{x_1+\xi \over 1+\xi}\ ,\qquad \qquad \!
{\vec{l}}_{\perp 1}
\ = \ {\vec{k}}_{\perp 1} + {1-x_1 \over 1+\xi}\,
\frac{{\vec{\Delta}}_\perp}{2} \ ,
\label{appn2}\\
y_{n+1}&=&{x_{n+1}-\xi \over 1+\xi}\ , \qquad \!
{\vec{l}}_{\perp n+1}
\ = \ {\vec{k}}_{\perp n+1} - {1+x_{n+1} \over 1+\xi}\,
\frac{{\vec{\Delta}}_\perp}{2} \ ,
\nonumber\\
y_i&=&{x_i\over 1+\xi}\ ,\qquad \qquad \
{\vec{l}}_{\perp i} \ = \ {\vec{k}}_{\perp i} - {x_i\over 1+\xi}\,
\frac{{\vec{\Delta}}_\perp}{2}
\qquad \qquad {\rm for}\ i = 2, \cdots ,n \ .
\nonumber
\end{eqnarray}
This satisfies $\sum^{n+1}_{i=1} y_i =1 $, $\sum^{n+1}_{i=1}
{\vec{l}}_{\perp i} ={\vec{0}}_\perp$ as required.  The off-diagonal
amplitude is non-zero in the domain $-\xi \le \bar{x} \le \xi$. There,
the formulae for the generalized form factors of the deeply virtual
Compton amplitude are
\begin{eqnarray}
\lefteqn{
{\sqrt{1-\xi^2}}\ H_{(n+1\to n-1)}(\bar{x},\xi,t)\,
-\, {\xi^2 \over {\sqrt{1-\xi^2}}}\
E_{(n+1\to n-1)}(\bar{x},\xi,t)
}
\label{t3app}\\
&=&
\sqrt{1-\xi}^{\, 3-n} \sqrt{1+\xi}^{\, 1-n}\,
\sum_{n, \lambda_i}
\int \prod_{i=1}^{n+1}
{{\rm d}x_{i}\, {\rm d}^2{\vec{k}}_{\perp i} \over 16\pi^3 }\
\nonumber\\
&& \qquad \qquad \rule{0pt}{3ex} {} \times
16\pi^3 \delta\left(1+\xi -\sum_{j=1}^{n+1} x_j\right) \,
\delta^{(2)}\left( \frac{{\vec{\Delta}}_\perp}{2} -
\sum_{j=1}^{n+1} {\vec{k}}_{\perp j}\right)
\nonumber\\
&& \qquad \qquad \rule{0pt}{3ex} {} \times
16\pi^3 \delta(x_{n+1}+x_{1}-2\xi)\,
\delta^{(2)}\left( {\vec{k}}_{\perp n+1} +
{\vec{k}}_{\perp 1} - {\vec{\Delta}}_\perp \right)
\nonumber\\[1ex]
&& \qquad \qquad \rule{0pt}{3ex} {} \times
\delta({\bar{x}}-x_1)\
\psi^{\uparrow\ *}_{(n-1)}
(x^\prime_i,{\vec{k}}^\prime_{\perp i},\lambda_i)
~ \psi^{\uparrow}_{(n+1)}(y_i,{\vec{l}}_{\perp i},\lambda_{i})\
\delta_{\lambda_1\, -\lambda_{n+1}} \ ,
\nonumber\\[1.5ex]
\lefteqn{
{1\over {\sqrt{1-\xi^2}}}\,
{(\Delta^1-{i} \Delta^2)\over 2M}\
E_{(n+1\to n-1)}(\bar{x},\xi,t)
}
\label{t3f2app}\\
&=&
\sqrt{1-\xi}^{\, 3-n} \sqrt{1+\xi}^{\, 1-n}\,
\sum_{n, \lambda_i}
\int \prod_{i=1}^{n+1}
{{\rm d}x_{i}\, {\rm d}^2{\vec{k}}_{\perp i} \over 16\pi^3 }\
\nonumber\\
&& \qquad \qquad \rule{0pt}{3ex} {} \times
16\pi^3 \delta\left(1+\xi -\sum_{j=1}^{n+1} x_j\right) \,
\delta^{(2)}\left( \frac{{\vec{\Delta}}_\perp}{2} -
\sum_{j=1}^{n+1} {\vec{k}}_{\perp j}\right)
\nonumber\\
&& \qquad \qquad \rule{0pt}{3ex} {} \times
16\pi^3 \delta(x_{n+1}+x_{1}-2\xi)\,
\delta^{(2)}\left( {\vec{k}}_{\perp n+1} +
{\vec{k}}_{\perp 1} - {\vec{\Delta}}_\perp \right)
\nonumber\\[1ex]
&& \qquad \qquad \rule{0pt}{3ex} {} \times
\delta({\bar{x}}-x_1)\
\psi^{\uparrow\ *}_{(n-1)}
(x^\prime_i,{\vec{k}}^\prime_{\perp i},\lambda_i)
~ \psi^{\downarrow}_{(n+1)}(y_i,{\vec{l}}_{\perp i},\lambda_{i})\
\delta_{\lambda_1\, -\lambda_{n+1}} \ .
\nonumber\end{eqnarray}

\section*{Acknowledgements}

We wish to thank  Paul Hoyer,  Bo-Qiang Ma, and Ivan Schmidt
for helpful suggestions.

\end{document}